\documentclass[%
 reprint,
superscriptaddress,
showkeys,
nofootinbib,
 amsmath,amssymb,
 aps,
pra,
floatfix,
]{revtex4-1}

\usepackage[colorinlistoftodos]{todonotes}

\usepackage[english]{babel}
\usepackage{hyperref}
\usepackage{algorithm}
\usepackage{algorithmic}
\usepackage{graphicx}
\usepackage{dcolumn}
\usepackage{bm}
\usepackage{color}
\graphicspath{ {images/} }
\usepackage{overpic}
\usepackage{mathrsfs}
\usepackage{algorithmic}
\usepackage{braket}

\newcommand{\sub}[1]{\textsubscript{#1}}
\newcommand{\super}[1]{\textsuperscript{#1}}
\usepackage{dcolumn}

\makeatletter
\newcommand{\algorithmicsync}{\textbf{sync}}
\newcommand{\algorithmicendsync}{\algorithmicend\ \algorithmicsync}
\renewenvironment{algorithmic}[1][0]{
\let\@item\ALC@item
  \newcommand{\ALC@lno}{%
\ifthenelse{\equal{\arabic{ALC@rem}}{0}}
{{\footnotesize \arabic{ALC@line}:}}{}%
}
\let\@listii\@listi
\let\@listiii\@listi
\let\@listiv\@listi
\let\@listv\@listi
\let\@listvi\@listi
\let\@listvii\@listi
  \newenvironment{ALC@g}{
    \begin{list}{\ALC@lno}{ \itemsep\z@ \itemindent\z@
    \listparindent\z@ \rightmargin\z@ 
    \topsep\z@ \partopsep\z@ \parskip\z@\parsep\z@
    \leftmargin 1em
    \addtolength{\ALC@tlm}{\leftmargin}
    }
  }
  {\end{list}}
  \newcommand{\ALC@it}{\addtocounter{ALC@line}{1}\addtocounter{ALC@rem}{1}\ifthenelse{\equal{\arabic{ALC@rem}}{#1}}{\setcounter{ALC@rem}{0}}{}\item}
  \newcommand{\ALC@com}[1]{\ifthenelse{\equal{##1}{default}}%
{}{\ \algorithmiccomment{##1}}}
  \newcommand{\REQUIRE}{\item[\algorithmicrequire]}
  
  \newcommand{\STATE}{\ALC@it}
  
  \newenvironment{ALC@if}{\begin{ALC@g}}{\end{ALC@g}}
  \newenvironment{ALC@for}{\begin{ALC@g}}{\end{ALC@g}}
  \newenvironment{ALC@whl}{\begin{ALC@g}}{\end{ALC@g}}
  \newenvironment{ALC@loop}{\begin{ALC@g}}{\end{ALC@g}}
  \newenvironment{ALC@rpt}{\begin{ALC@g}}{\end{ALC@g}}
  \newenvironment{ALC@syn}{\begin{ALC@g}}{\end{ALC@g}}
  \renewcommand{\\}{\@centercr}

  \newcommand{\IF}[2][default]{\ALC@it\algorithmicif\ ##2\ \algorithmicthen%
\ALC@com{##1}\begin{ALC@if}}
  \newcommand{\ELSE}[1][default]{\end{ALC@if}\ALC@it\algorithmicelse%
\ALC@com{##1}\begin{ALC@if}}
  \newcommand{\ELSIF}[2][default]%
{\end{ALC@if}\ALC@it\algorithmicelsif\ ##2\ \algorithmicthen%
\ALC@com{##1}\begin{ALC@if}}
  \newcommand{\FOR}[2][default]{\ALC@it\algorithmicfor\ ##2\ \algorithmicdo%
\ALC@com{##1}\begin{ALC@for}}
  \newcommand{\FORALL}[2][default]{\ALC@it\algorithmicforall\ ##2\ %
\algorithmicdo%
\ALC@com{##1}\begin{ALC@for}}
  \newcommand{\WHILE}[2][default]{\ALC@it\algorithmicwhile\ ##2\ %
\algorithmicdo%
\ALC@com{##1}\begin{ALC@whl}}
  \newcommand{\LOOP}[1][default]{\ALC@it\algorithmicloop%
\ALC@com{##1}\begin{ALC@loop}}
  \newcommand{\REPEAT}[1][default]{\ALC@it\algorithmicrepeat%
\ALC@com{##1}\begin{ALC@rpt}}
  \newcommand{\UNTIL}[1]{\end{ALC@rpt}\ALC@it\algorithmicuntil\ ##1}
\newcommand{\SYNC}[2][default]{\ALC@it\algorithmicsync\ ##2\ 
    \ALC@com{##1}\begin{ALC@syn}} 
  \ifthenelse{\boolean{ALC@noend}}{
    \newcommand{\ENDIF}{\end{ALC@if}}
    \newcommand{\ENDFOR}{\end{ALC@for}}
    \newcommand{\ENDWHILE}{\end{ALC@whl}}
    \newcommand{\ENDLOOP}{\end{ALC@loop}}
    \newcommand{\ENDSYNC}{\end{ALC@syn}} 
  }{
    \newcommand{\ENDIF}{\end{ALC@if}\ALC@it\algorithmicendif}
    \newcommand{\ENDFOR}{\end{ALC@for}\ALC@it\algorithmicendfor}
    \newcommand{\ENDWHILE}{\end{ALC@whl}\ALC@it\algorithmicendwhile}
    \newcommand{\ENDLOOP}{\end{ALC@loop}\ALC@it\algorithmicendloop}
    \newcommand{\ENDSYNC}{\end{ALC@syn}\ALC@it\algorithmicendsync}
  } 
  \renewcommand{\@toodeep}{}
  \begin{list}{\ALC@lno}{\setcounter{ALC@line}{0}\setcounter{ALC@rem}{0}%
    \itemsep\z@ \itemindent\z@ \listparindent\z@%
    \partopsep\z@ \parskip\z@ \parsep\z@%
    \labelsep 0.5em \topsep 0.2em%
\ifthenelse{\equal{#1}{0}}
  {\labelwidth 0.5em }
  {\labelwidth  1.2em }
\leftmargin\labelwidth \addtolength{\leftmargin}{\labelsep}
    \ALC@tlm\labelsep
  }
}
{\end{list}}
\makeatother

\definecolor{blue(pigment)}{rgb}{0.2, 0.2, 0.6}
\definecolor{darkerblue}{rgb}{0.0, 0.0, 0.4}
\definecolor{darkblue}{rgb}{0.0,0.0,0.5}
\definecolor{darkgreen}{rgb}{0.0,0.4,0.0}
\hypersetup{
    colorlinks,
    linkcolor=blue(pigment),
    citecolor=darkgreen,
    urlcolor=darkblue
}

\usepackage{adjustbox}
\usepackage{array}

\newcolumntype{R}[2]{%
    >{\adjustbox{angle=#1,lap=\width-(#2)}\bgroup}%
    l%
    <{\egroup}%
}

\usepackage{multirow}

\begin{document}


\title{Optically Controlled Entangling Gates in Randomly Doped Silicon}

\author{Eleanor Crane}
 \email{e.crane@ucl.ac.uk}
\affiliation{London Centre for Nanotechnology, University College London, Gower Street, London WC1E 6BT, United Kingdom}
\author{Thomas Crane}
\affiliation{Laboratoire SPHERE, Universit\'e Paris Diderot, 5 Rue Thomas Mann, Paris 75013, France}
\author{Alexander Schuckert}
\affiliation{Department  of  Physics, Technical  University  of  Munich,  85748  Garching,  Germany}
\author{Nguyen H. Le}
\affiliation{Advanced Technology Institute and Department of Physics, University of Surrey, Guildford GU2 7XH, United Kingdom}
\author{Andrew J. Fisher}
\affiliation{London Centre for Nanotechnology, University College London, Gower Street, London WC1E 6BT, United Kingdom}
\affiliation{Department of Physics \& Astronomy, University College London, Gower St, London, WC1E 6BT, United Kingdom}

\begin{abstract}
Randomly-doped silicon has many competitive advantages for quantum computation; not only is it fast to fabricate but it could naturally contain high numbers of qubits and logic gates as a function of doping densities. We determine the densities of entangling gates in randomly doped silicon comprising two different dopant species. First, we define conditions and plot maps of the relative locations of the dopants necessary for them to form exchange interaction mediated entangling gates. Second, using nearest neighbour Poisson point process theory, we calculate the doping densities necessary for maximal densities of single and dual-species gates. We find agreement of our results with a Monte Carlo simulation, for which we present the algorithms, which handles multiple donor structures and scales optimally with the number of dopants and use it to extract donor structures not  captured by our Poisson point process theory. Third, using the moving average cluster expansion technique, we make predictions for a proof of principle experiment demonstrating the control of one species by the orbital excitation of another. These combined approaches to density optimization in random distributions may be useful for other condensed matter systems as well as applications outside physics.
\end{abstract}

\maketitle

\section{Introduction}

The ability to perform coherent quantum operations on large collections of quantum bits (qubits) could lead not only superior scaling to classical machines on date-based algorithms such as factoring~\cite{shor_algorithms_1994}, unstructured search~\cite{grover_fast_1996} and machine learning ~\cite{biamonte_quantum_2017} but may also lead to a better understanding of strongly correlated systems through Feynman's idea of quantum simulation~\cite{feynman_simulating_1982, king_observation_2018, braumuller_analog_2017, lamata_analog_2018}. Initial successes in the realization of these applications in the last decades, mainly on AMO platforms such as trapped ions~\cite{haffner_quantum_2008,zhang_observation_2017} and ultracold atoms in optical lattices~\cite{greiner_quantum_2002,gross_quantum_2017,mazurenko_cold-atom_2017,bernien_probing_2017}, have also shown how hard it is to reach a regime where quantum computers are large enough to outperform their classical counterparts either due to limitations in the qubit addressability, interactions or sheer quantity. This scalability challenge could be more easily overcome in solid-state realizations, where a wide range of qubits have been proposed including Majorana fermions in nanowires~\cite{sau_viewpoint_2017,lahtinen_short_2017}, superconducting qubits~\cite{devoret_superconducting_2004, barends_superconducting_2014, kelly_state_2015}, nitrogen-vacancy centers in diamond~\cite{childress_diamond_2013}, quantum dots~\cite{mi_circuit_2017}, and donor impurities in silicon~\cite{hill_surface_2015,morley_review_2014}.

A realization in doped silicon, the most important material in the electronics industry today, could point the way forward for the widespread introduction of quantum computation.
\begin{figure}[H]
\includegraphics{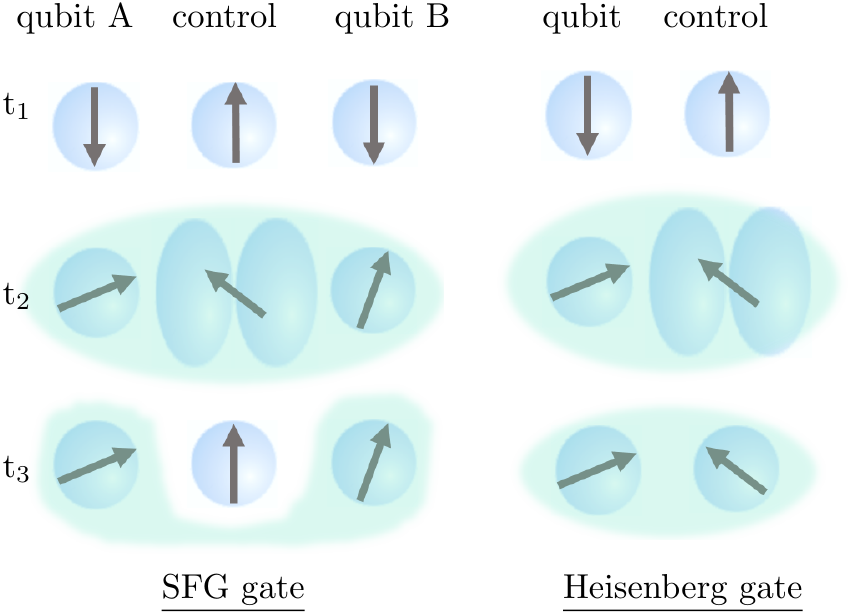}
\caption{\label{fig_illustration}\textbf{Optically controlled interaction exchange energy entangling gates in donors in silicon.} At time $t_1$, all donors are non-interacting in their ground state; at time $t_2$, the controls are excited to the a higher lying and physically extended orbital state, entangling themselves via Heisenberg exchange with the neighbouring qubits of a different donor species; finally at $t_3$ the control donor electron falls back to the ground state.  In the SFG scheme (left) the excited controls mediate entangling interactions between pairs of qubits and the control is then removed from the entangled state on de-excitation, while in the Heisenberg scheme (right) the entanglement is produced between a qubit and the control.}
\end{figure}
\newpage
Donor impurities have several spin 1/2 degrees of freedom which provide natural Hilbert spaces to encode qubits~\cite{divincenzo_quantum_1998}. The nuclear spins of ${}^{31}$P~\cite{kane_silicon-based_1998,morton_solid-state_2008} and the electron spin of the valence electron of several species have some of the longest decoherence times of any qubit realizations ranging from hours to several days~\cite{pla_high-fidelity_2013} especially when benefiting from isotopically pure ${}^{28}$Si and/ or field-insensitive “clock transitions”~\cite{saeedi_room-temperature_2013, saeedi_optical_2014}. Initialisation and readout~\cite{morello_single-shot_2010} and single-qubit operations have been demonstrated on single electron spins ~\cite{pla_high-fidelity_2013}, via metallic microwave strip lines~\cite{fuechsle_single-atom_2012} and donor-bound exciton spectroscopy~\cite{lo_hybrid_2015, saeedi_donorbound_2015}. However, in addition to initialisation, readout and single-qubit operations, a two-qubit entangling gate is required for universal quantum computation~\cite{lloyd_universal_1996}.

Several proposals to realise entangling gates in silicon exploit the exchange interactions between ground-state electronic states; these interactions are strongly oscillatory because of interference between the different valley components of donor wavefunction~\cite{koiller_exchange_2001} and typically require atomic-scale precision in the positioning of the donors~\cite{kane_silicon-based_1998}. This needs specialised lithography based on selective removal of hydrogen atoms in the scanning tunnelling microscope~\cite{hersam_silicon-based_2000} and subsequent implantation of donor species ~\cite{obrien_scanning_2002}, a delicate and resource-intensive technique\cite{obrien_scanning_2002, fuechsle_single-atom_2012, broome_two-electron_2018}.   Schemes based on electric-dipole interactions have larger donor spacings and tolerate less stringent fabrication requirements~\cite{tosi_silicon_2017}.  However, a scheme capable of inducing entangling interactions with truly random placement of donors would allow relatively seamless integration of the fabrication into the standard widespread silicon processing industry.

In this paper, we discuss a family of entangling gates, sketched in Fig.~\ref{fig_illustration}, that make use of physically localised or extended orbital excited states to control the interactions between donor qubits in a disordered ensemble~\cite{stoneham_optically_2003}. These orbital excited states can be produced by excitation with terahertz radiation~\cite{greenland_coherent_2010}.  The random locations of the donors provides inhomogeneous broadening due to differing local magnetic field provided by hyperfine spins which could enable selective addressing of sub-ensembles with a specific frequency of terahertz radiation~\cite{stoneham_room-temperature_2009}, and the spread of frequencies could be further increased by placing the sample in a field gradient, leading to each donor having a different transition frequency~\cite{stoneham_room-temperature_2009}. The excited states produced (for example, 2p orbitals) have a spatial structure that depends on the axis of polarisation of the laser, thus providing another tool for spatial selectivity within local configurations. Important steps have been made towards the realization of exchange interaction-based gates in donors in silicon. The disentanglement of the control particle from the two entangled qubits was shown to be feasible after the gate operation~\cite{rodriquez_avoiding_2004}. Coherent control of the valence electron orbital states of dopants such as phosphorous, bismuth, antimony and arsenic has been demonstrated with terahertz light, using a Free Electron Laser tuned to the low tens of meV~\cite{greenland_coherent_2010}. Furthermore, initialisation and readout can be done optically, using donor-bound exciton spectroscopy in the terahertz frequencies which was demonstrated experimentally with phosphorous spins in silicon~\cite{morse_zero-field_2018}. Until now it has remained unclear what densities of `viable' entangling gates could be reached and even what the requirements are for a configuration to be `viable'.

The present paper deals only with the special case of homogeneous distributions, either throughout a 3D sample or on a 2D plane.  These correspond to the ideal limits of a uniformly doped bulk sample or a perfect delta-layer respectively; we also consider the case where two parallel ideal delta-layers are implanted with different species.  The general case of a spatially varying density (hence allowing both for graded doping of a bulk material and for the inevitable broadening of real delta-layers) is treated in a companion paper~\cite{paper2}; in that paper a heuristic is also given which enables fast estimates of the densities of viable clusters over a wide density range.

This paper is organized as follows. First, we estimate appropriate conditions on the spacing of the dopants required for them to form a viable entangling gate. Second, optimal doping densities to produce the highest densities of entangling gates are calculated using a Monte-Carlo simulation and Poisson point process theory. Third, we propose a proof-of-principle experiment to show control over the exchange interaction between donors at the optimal doping densities and predict its results with the Moving Average Cluster Expansion (MACE) method~\cite{hazzard_many-body_2014}.  Our numerical results apply to the case of Group V donors in silicon; however, our methods apply to any material where particular configurations of multiple impurities are required, and the conditions on the configurations needed can be expressed in terms of distances between different species.

\section{Geometric conditions on optically controlled entangling gate configurations}
First, we define conditions on the distances between dopants to form an SFG or Heisenberg entangling gate by comparing energy scales corresponding to the lifetime of the excited state~\cite{vinh_silicon_2008,litvinenko_coherent_2015}, the strength of the entangling exchange interaction and the requirement of qubits to be isolated from each other when not part of the entangling operation. Based on these conditions we then calculate maps and line scans of the interaction exchange energy from a theory including multi-valley interference~\cite{wu_excited_2018} to delimit areas around donors in which entangling gate operation is possible.

\subsection{Conditions on the exchange interaction energy}

A phase gate (which can be combined with Hadamard gates to result in a Controlled NOT gate)~\cite{levy_universal_2002} can be implemented by the sequence
\begin{equation}
\hat{\textit{u}}_{phase} = \hat{r_2^z}\big(-\frac{\pi}{2}\big)\hat{r_1^z}\big(\frac{\pi}{2}\big)
\hat{e}_{12}\big(\frac{\pi}{2}\big)\hat{r_1^z}\big(\pi\big)\hat{e}_{12}\big(\frac{\pi}{2}\big),
\end{equation}
where the single qubit operations are given by the Zeeman rotations $\hat{r}_i^\alpha = e^{-i\theta \hat{H}_i^\alpha / g B^\alpha}$. The two qubit operation is given by $\hat{e}_{12}(\frac{\pi}{2}) = e^{-i\frac{\pi}{2} \hat{H}_{12} / J}$, where $\hat{H}_{12}= J \hat{\mathbf{S}}_1 \hat{\mathbf{S}}_2$ is the Heisenberg Hamiltonian between spins $\mathbf{S}_1$ and $\mathbf{S}_2$. Equating this with the time evolution operator $\hat{U}=e^{-\frac{i}{\hbar} t \hat{H}_{12}}$ leads to the condition
\begin{equation}
J = \frac{h}{4t} \quad\mathrm{or}\quad t=\frac{h}{4J}
\label{1_Eq_condJ}
\end{equation}
for a successful phase gate operation.

The exchange interaction constant $J$ arises from overlap of spratially separated wavefunctions.  The value of $J$ (or $t$) must be controlled such that Eq.(\ref{1_Eq_condJ}) is fulfilled, without perturbing interactions between the gate dopants and other donors.  In silicon this is especially challenging as the exchange interaction between valence dopant electrons not only decays exponentially with distance but also oscillates with a period of the order of the lattice spacing due to intervalley-interference~\cite{keyes_challenges_2005}. The lifetimes $T_ {\mathrm{dec}}$ of the donor excited states  are limited by phonon-mediated decay, and for phosphorous and arsenic they have been measured as $T_ {\mathrm{dec}}=200\,\mathrm{ps}$\cite{vinh_silicon_2008}, which limits the gate operation time to $t<T_ {\mathrm{dec}}$; hence, we define an energy scale 
\begin{equation}
J_\mathrm{dec}=\frac{h}{4 T_\mathrm{dec}}.
\label{eq_Jdec}
\end{equation}
which sets the minimum exchange energy needed for being able to perform a successful 2-qubit gate.

In the rest of this paper we will consider two group-V donor species: the shallow donor phosphorous (Si:P) as the controls (also referred to as c in the general case), whose excited state wavefunctions (P$_{2p+-}$) mediate the interactions between the deeper arsenic (Si:As) ground state (As$_{1s}$) spins as the readout qubits (also referred to as r if not specifically considering an implementation relying on using this particular species).

\subsubsection{SFG entangling gate}
\begin{figure}
\includegraphics{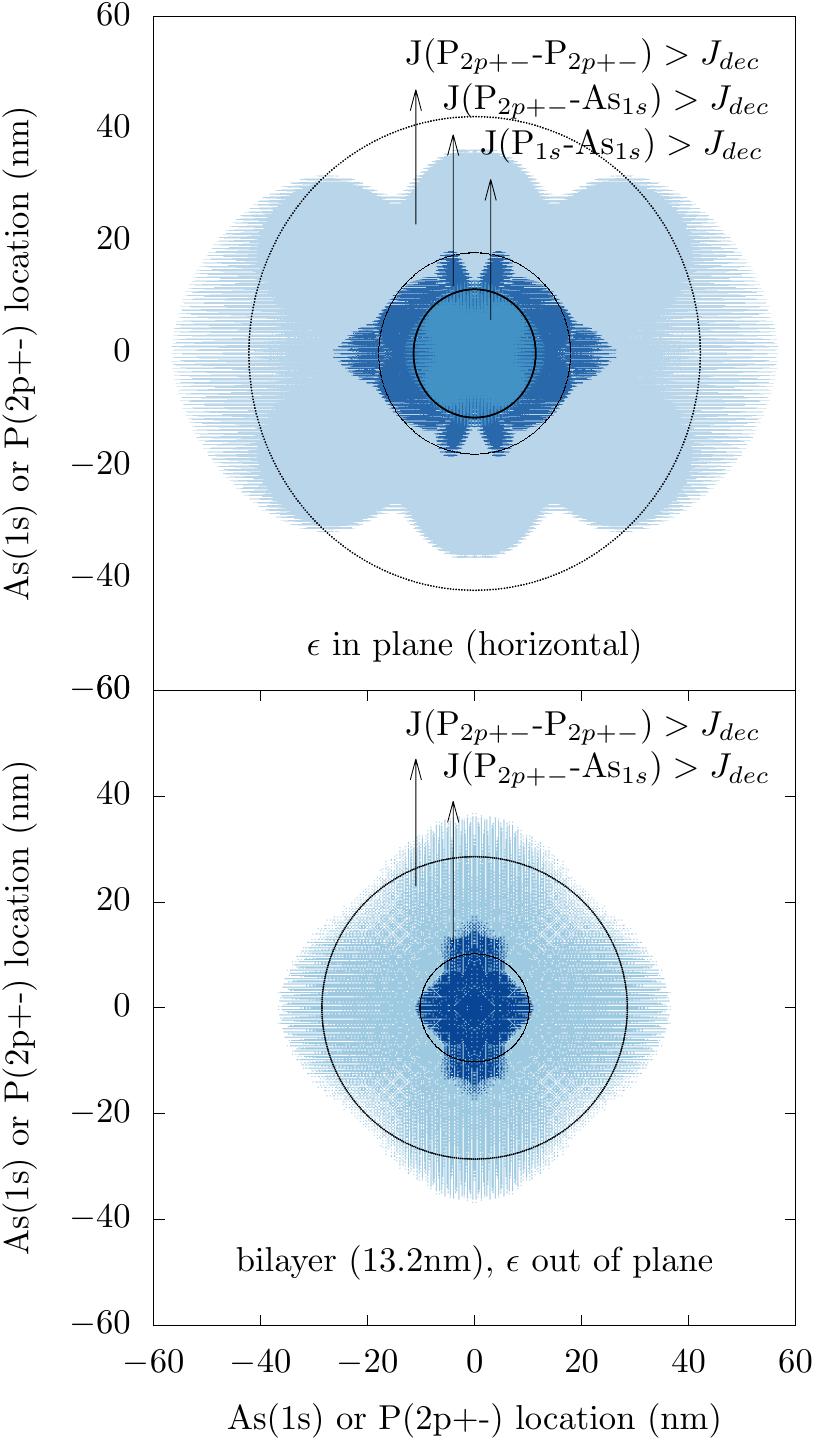}
\caption{\label{fig_jex_maps}\textbf{Interaction exchange energy limits} in the case where arsenic and phosphorous are doped in the same layer and in the case where arsenic is doped in a parallel layer 13.2nm away to avoid ground state interaction. Circles represent the approximations. $\epsilon$ is the laser light polarisation axis.}
\end{figure}
We assume that only the control donors are excited to a higher-lying orbital state which we will choose to be the 2p excited state; the readout donors remain in the ground state. The following conditions then need to be met for successful gate operation:
\begin{description}
\item[Readouts] Interaction between readouts should be smaller than the decoherence interaction exchange energy defined in Eq.(\ref{eq_Jdec}), ie.  $J$(r$_{1s}$-r$_{1s})(r) < J_\mathrm{dec}$. We will denote the distance at which these two scales are equal as $R_{rr}$.
\item[Readout-Control] If the control is in the ground (1s) state, the interactions between readout and control should also be small: $J$(c$_{1s}$-r$_{1s})(r) < J_\mathrm{dec}$, with equality at $R_{min}$. Furthermore, if the control is in the excited state, control and readout should interact, therefore $J$(c$_{2p}$-r$_{1s})(r) > J_\mathrm{dec}$ with equality at $R_{max}$.
\item[Controls] If two neighbouring controls are excited to the $2p$ their interaction should not dominate the process, i.e. $J$(c$_{2p}$-c$_{2p})(r) < J_\mathrm{dec}$ with equality at $R_{cc}$.
\end{description}
The distances introduced above are shown in Fig.~\ref{fig_jex_maps}.

Please note that in the above considerations we did not need to specify the number of readouts per control. In fact, although originally devised for two readouts~\cite{stoneham_optically_2003}, we will leave this number general.

\subsubsection{Heisenberg entangling gate}
There are two ways of implementing this gate. In the first, which we refer to as `excited-ground Heisenberg gate', one donor is in the excited and the other in the ground state; this configuration is the same as the SFG gate with just one readout. In the second case, the `excited-excited Heisenberg gate', both donors are excited. Therefore this can be a single-species gate composed of two Si:P electrons, for example.

In the excited-ground case, the conditions are the same as for the SFG gate, except that only one readout should interact strongly with a control, and not two.

For the excited-excited Heisenberg gate the dominant process should be the 2p+- state of both controls interacting with each other: J(c$_{2p}$-c$_{2p})(r) > J_\mathrm{dec}$ with equality at $R_{max}=R_{cc}$ and the entire gate should also be isolated from other donors by this distance. The interaction of ground-state controls in $1s$ should be small: J(c$_{1s}$-c$_{1s})(r) < J_\mathrm{dec}$ with equality at $R'_{min}$.

Having defined the conditions that viable entangling gate configurations need to fulfil, we can now find the corresponding distances by calculating maps of the exchange interaction between Si:P and Si:As.

\subsection{Constraints on dopant separations from conditions on exchange interaction}

The exchange interaction energy calculated using the Heitler-London approximation~\cite{koiller_exchange_2001,wu_exchange_2008} taking into account the multi-valley coupled wavefunction and central cell correction can be found in Appendix \ref{app_jex_calc}. The maps in Fig. \ref{fig_jex_maps} show the regions in which $J > J_{\mathrm{dec}}$, for $J$ connecting different types of dopant species in various states.

For information to not remain localised in the quantum computer, it is essential that qubits can communicate with two or more entangling gates. This has the implication that the maximum separation between controls and qubits is larger than half of the distance separating two controls / gates. This is the case, as can be seen in Fig. \ref{fig_jex_maps}.

To simplify the subsequent treatment of the dopant distributions, we approximate the interaction zones outlined by $J_{\mathrm{dec}}$ as circles or spheres; to do this, we determine discs whose areas are equal to those of the interaction zones. The disc radii corresponding to the monolayer doping in plane laser polarisation setup and the bilayer out of plane laser polarisation described in the next paragraph are displayed in Table. \ref{tab_cond}.

\begin{table}
\centering
\begin{tabular}{c|c|c|c}
Polarisation & Interaction =$J_{\mathrm{dec}}$ & Name & Distance \\
 axis of light &  &  & $(nm)$\\
\hline
\\[-1em]
\multirow{5}{*}{In plane} & J(As$_{1s}$-As$_{1s}$) & $R_{rr}$ & 11.0 \\
\\[-1em]
& J(P$_{1s}$-As$_{1s}$) & $R_{min}$  & 11.4\\
\\[-1em]
& J(P$_{2p}$-As$_{1s}$)& $R_{max}$ & 17.9\\
\\[-1em]
 & J(P$_{2p}$-P$_{2p}$) & $R_{cc}$ & 42.2\\
\\[-1em]
 & J(P$_{1s}$-P$_{1s}$) & $R_{min}'$ & 11.8\\
\hline
\\[-1em]
\multirow{2}{*}{Out of plane}  & J*(P$_{2p}$-As$_{1s})$ & $R_{max}$* & 10.2\\
\\[-1em]
 & J(P$_{2p}$-P$_{2p}$) & $R_{cc}$ & 28.5\\
\hline
\end{tabular}
\caption{\textbf{Values for the minimum/maximum radius between dopants occurring when their interaction exchange energy is equal to $J_{\mathrm{dec}}$}, defined in Eq. 3, plotted in Fig. \ref{fig_jex_maps} for the operation of the optical entangling gate within the decay time.  *Bilayer case - calculated for readouts (Si:As) in a plane 13.2nm from and parallel to the control (Si:P) plane.}
\label{tab_cond}
\end{table}

\subsubsection{Bilayers}\label{sec:duolayers}
For two-species gates, we also consider separating the controls and the readouts into two separate parallel planes, sufficiently distant that the ground states do not interact on the timescale of the gate operation but sufficiently close that the qubits can interact with the excited state of the controls. For this to be optimised in a randomly doped sample, the readout layer must be at a distance from the control layer which maximises exposure to the area in which $J$(P$_{2p}$-As$_{1s})<$ $J_{\mathrm{dec}}$ and minimum exposure to the area in which $J$(P$_{1s}$-As$_{1s})>$ $J_{\mathrm{dec}}$, dark blue on Fig.~\ref{fig_jex_maps}. The small periodic oscillation put apart, the  ground state interaction exchange energies are equal $J_{\mathrm{dec}}$ in a spherically symmetric way, and can be modelled as a sphere. As is clear from Fig. \ref{fig_ppp}b), the optimal distance $d$ separating the two layers should be equal to $R_{min}$. For the 1s states of Si:As and Si:P this is 13.2\,nm. Note that the areas which the qubit can occupy in its own layer correspond to a disc of inner radius 0 and outer radius $\sqrt{R_{max}^2\:-\:d^2}$, as can be seen in Fig.~ \ref{fig_ppp}b). By polarising the light out of plane the controls can pack closer together and thus make more qubits viable thereby increasing the density of entangling gates.

\section{Maximising the entangling gate density}

In a randomly doped sample, the only free parameters are the doping densities. In this section, we establish the optimal doping densities of both species such that the entangling gate density is maximal. Phosphorous atoms have a random spatial arrangement on the silicon surface resulting from exposure to phosphine gas~\cite{trappmann_observation_1997}. The atoms cannot land in exactly the same location, thus their arrangement would correspond at best to a hard sphere Poisson point process, however we make the approximation that they correspond to a pure Poisson point process. Many methods, including analytical and simulation-based approaches, are available to treat Poisson point processes: Monte-Carlo simulations can treat problems with complex geometries and have the advantage of being flexible, but must be coded efficiently to scale well. Analytical approaches can supply results in closed form but rely on particular assumptions and are valid only if these apply.  Analytical nearest-neighbour methods have been used to study distributions of $m^{th}$ nearest neighbours~\cite{torquato_nearest-neighbour_1989}, isolated pairs of points~\cite{pickard_isolated_1982} and the probability of occurrence of reflexive nearest neighbours~\cite{cox_reflexive_1981}\cite{dacey_proportion_2010}.

Here we employ a Monte Carlo simulation to count viable configurations (defined in the previous section). It scales linearly with the number of dopants, i.e. as $\mathcal{O}(n_c + n_r)$, where $n_c$ is the number of control dopants and $n_r$ is the number of readout dopants.  This allows large samples (up to a few million dopants per run) to be used routinely, leading to lower statistical errors. The algorithm used is described in Appendix \ref{app_simu}. We find agreement of these results with an analytical solution for a Poisson point process in both two and three spatial dimensions, using nearest-neighbour methods to determine the probability of occurrence of a nearest event within a given radius from a point chosen at random~\cite{bahcall_distribution_1981}. We treat the two dopant species as separate independent sets of events occupying the same volume, with different densities.

First, we study the control dopant distribution and calculate the control density which gives the highest density of points separated by at least $R_{cc}$; we call the total density of such controls the \textit{viable} control density. 

Second, we find the cumulative probability distribution function (CPDF) describing the probability that a control chosen at random (we make the approximation that the isolated control dopants are randomly distributed) is surrounded by a configuration of readouts fulfilling the distance conditions in Table. \ref{tab_cond}; we seek the optimal readout density where the greatest number of controls have both 1st and 2nd nearest readout neighbours that are situated in the `viability shell' between $R_{min}$ and $R_{max}$, are further than $R_{rr}$ from each other, and are also further than $R_{rr}$ from any other neighbour. We obtain a CPDF which depends on $R_{min}$ and $R_{max}$ and $R_{rr}$.

Third, we find the optimal density of viable entangling gate configurations by combining the two previous results, by making the approximation that both probabilities relative to the control and readout are independent.

\begin{figure*}
\includegraphics{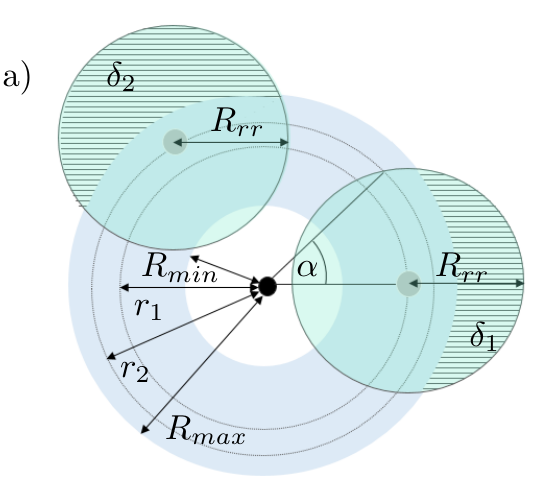}\quad \includegraphics{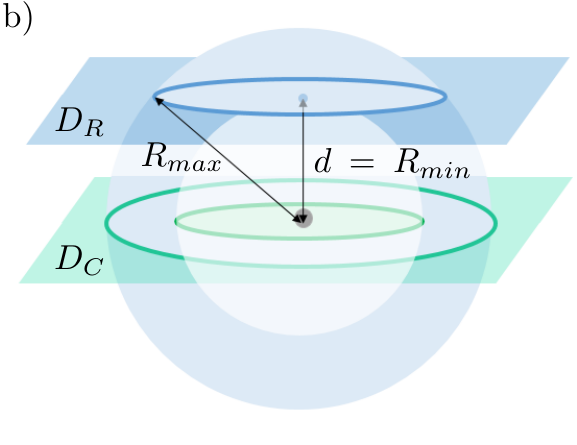}
\caption{\textbf{Schematic representations of geometric considerations for readout dopants surrounding the control dopant.} a) The first and second nearest neighbours of a random point chosen in the readout distribution must be separated by $R_{rr}$, which is calculated using the exclusion angle $\alpha$ and the hashed areas referred to as $\delta$. b) Separating the 2D distribution of control dopants $D_c$ and readout dopants $D_r$ into parallel bilayers enables the determination of a distance at which the viability sphere has the most area exposed to the readout distribution. The optimal separation between the layers is d = $R_{min}$, as can be worked out from applying Pythagoras.}
\label{fig_ppp}
\end{figure*}

We find slightly higher entangling gate densities for the bilayer configuration described in Fig. \ref{fig_ppp}, for which each species is contained in a 2D layer separated by a small distance.  In the Monte Carlo simulation, we are also able to identify configurations with more than two readouts within the viability shell surrounding the same control. These could be useful for implementing multi-qubit entangling gates such as the Toffoli gate~\cite{monz_realisation_2009}. 

\subsection{Control doping density: maximising the density of events isolated by a fixed radius}

\begin{figure}
\includegraphics{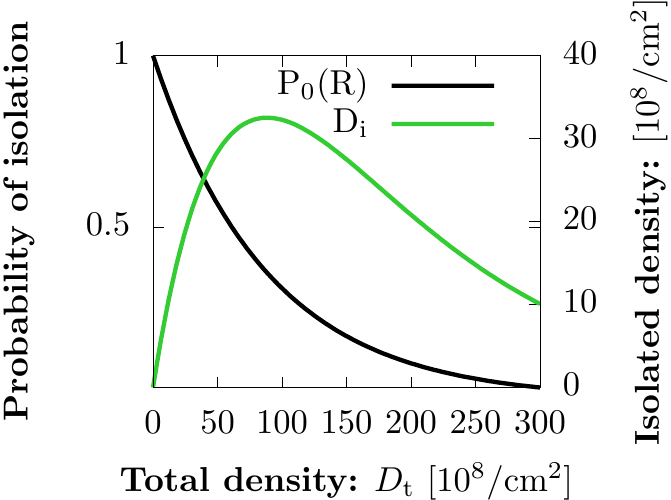}
\caption{\textbf{$\textbf{P}_{\textbf{0}}\textbf{(R}\textbf{)}$, probability for a point to be isolated by a fixed radius} (no clear maximum to optimise for) and \textbf{D\textsubscript{i}, density of points isolated by the same fixed radius} (containing a maximum which it is possible to optimise for), as \textbf{functions of the total density, D\textsubscript{t}.}}
\label{graph_isolation}
\end{figure}

We seek to maximise the density of events isolated by at least a fixed radius $R_{cc}$. The probability of finding an event in a spherical shell of thickness dr at a distance $r$ from a randomly chosen event in the distribution of density $D_t$ is $\frac{\mathrm{d} V(r)}{\mathrm{d}r}D_t\,dr$, where $V(r)$ is the volume of the sphere of radius $r$~\cite{scott_nearest_1989, moltchanov_distance_2012}. The probability that there are no events closer to the randomly chosen event than $r$ is then $P_0(r) = e^{-V(r) D_t}$, from~\cite{bahcall_distribution_1981}\cite{scott_nearest_1989}. Optimising this probability leads to setting the total density of events to 1 (such that there is no chance of finding any other event within $r$), as can be seen on Fig.~\ref{graph_isolation}. However, optimising the density of points $D_{i}$ isolated by exactly $R$ from their neighbours does present a maximum (see Fig.~\ref{graph_isolation}).
\begin{equation}
D_{i}\equiv D_t \: P_0(R) = D_{t}e^{-V(R)D_{t}},
\label{eq_Di}
\end{equation}
which is maximum when
\begin{align}
D_{t}(D_{i\:max}) = \frac{1}{V(R)}.
\label{eq_isolation_density}
\end{align}
At this density, the fraction of points isolated by $R$ is $1/e$:
\begin{align}
D_{i\:max} = D_{t}\:e^{- 1}.
\label{vdens}
\end{align}

The values for the total and isolated control densities corresponding to the control radii (Table.~\ref{tab_cond}) are displayed in Table.~\ref{tab_control_res}.
 
\begin{table}
\centering
\begin{tabular}{c|c|c|c|}
\multicolumn{1}{c|}{} & 3D ($10^{15}$/cm$^3$) & \multicolumn{2}{c|}{2D ($10^{10}$/cm$^2$)}\\ 
\cline{2-4}
 \multicolumn{1}{c|}{} & \multicolumn{2}{c|}{R = 42.2 nm} & R* = 28.5 nm\\
\hline 
Total density & 3.18 & 1.79 & 3.91\\ 
\hline 
Isolated density & 1.17 & 0.66 & 1.44  \\ 
\hline 
Isolated fraction & \multicolumn{3}{c|}{1/e} \\ 
\hline 
\end{tabular} 
\caption{\label{tab_control_res} Total doping densities giving maximum number of dopants in 3D, and in 2D isolated by R. *radius corresponding to laser light polarisation out of plane.}
\end{table}

\subsection{Readout doping density}

Let $S(r)=\frac{\mathrm{d}V(r)}{\mathrm{d}r}$ be the surface area of the $n$-dimensional sphere of radius $r$ (i.e. $S(r)=2 \pi r$ in 2D and $S(r)=4 \pi r^2$ in 3D).

If the readout density is $D_r$, the probability of finding the first and second nearest neighbours between $R_{min}$ and $R_{max}$ is
\begin{equation}
\int_{R_{min}}^{R_{max}}dr_1\int_{r_1}^{R_{max}}dr_2\:S(r_{1}) S(r_{2}) D_{r}^2\; e^{-V(R_{max})D_{r}}.
\label{s1s2}
\end{equation}

However, the number of viable configurations is reduced by the additional requirement that the second nearest neighbour must be at least $R_{rr}$ from the first. If $r_2 < (r_1 + R_{rr})$, the sphere of radius $r_2$ defining the viable positions for the second nearest neighbour must therefore have a spherical cap removed from it (see Fig.~\ref{fig_ppp}) subtending an angle $\alpha = \cos^{-1}\big( \frac{r_{1}^2 + r_{2}^2 - R_{rr}^2}{2 r_{1} r_{2}}\big)$.
This gives rise to a new surface of smaller area $\tilde{S}(r_2)=2 (\pi-\alpha) r_2$ in 2D and $\tilde{S}(r_2)=4 (\pi-\alpha) r_2^2$ in 3D.
In the bilayer case, $R_{min} = 0$ which further constrains $\alpha$ to be $\pi$ if $r_1 < (R_{rr} - R_{max})$, ie. if $\alpha$ is complex.

Finally, cases where further readout donors lie outside the sphere of radius $R_{max}$ but within a volume $\delta$ defined as being within radius $R_{rr}$ of the first or second nearest neighbour must be excluded; this corresponds to reducing the probability by a factor
\begin{equation}
e^{-\delta\,D_{r}}. \label{delt}
\end{equation}
In the SFG case, if $\delta_1(r_1)$ and $\delta_2(r_2)$ are the volumes of the spheres of radius $R_{rr}$ centred on the first and second nearest neighbours which lie outside the sphere of volume $R_{max}$, and $\delta_{ov}(r_1,r_2,\theta)$ is the overlap between these two volumes where $\theta$ is the angle between the readouts, $\delta_{SFG}=\delta_1(r_1)+\delta_2(r_2)-\delta_{ov}(r_1,r_2,\theta)$. In both other gate types, $\delta=\delta_1(r_1)$. This three circle overlap configuration and area calculation has been made into a Wolfram Demonstrations project~\cite{crane_wolfram_2019} and the functions corresponding to the overlap of two and three circles case was taken from~\cite{fewell_area_2006}. 

We can now construct the total densities of viable configurations for our three types of entangling gate, for homogeneous doping with a control density $D_c$ and a readout density $D_r$.  
\begin{itemize}
\item[(i)] The density of SFG gate configurations is the product of the viable control density with the probability for a control to be surrounded by a successful configuration of readouts (the product of the CPDFs defined in equations \ref{s1s2} and \ref{delt}):
\begin{align}\label{eq_Dsfg}
&\qquad\qquad D_{\mathrm{sfg}} (D_c,D_r) = D_{c}\:e^{-  V(R_{cc}) \: D_{c}} \times\\
&\int_{R_{min}}^{R_{max}}dr_1\int_{r_1}^{R_{max}}dr_2 S(r_1) \tilde{S}(r_2) D_{r}^2 e^{-(V(R_{max})+ \delta_{SFG})D_r}\notag
\end{align}

\item[(ii)]The density of two-species excited-ground Heisenberg entangling gates is:
\begin{align}\label{eq_Dheisex-gd}
&D_{\mathrm{Heis.\:ex-gd}} (D_c,D_r) = D_c\:e^{-  V(R_{cc}) \: D_c} \times  \\
&\quad\int_{R_{min}}^{R_{max}} dr S(r)  D_r^2 e^{-(V(R_{max})+ \delta)D_r};\notag
\end{align}

\item[(iii)]The density of single species excited-excited Heisenberg entangling gates is
\begin{align}
D_{\mathrm{Heis.\:ex-ex}} (D) &= D \int_{R_{min}'}^{R_{cc}} dr S(r)  D^2 e^{-(V(R_{cc})+ \delta)D}.
\label{eq_Dheisex-ex}
\end{align}
\end{itemize}

\subsection{Results}

\begin{figure}
\includegraphics{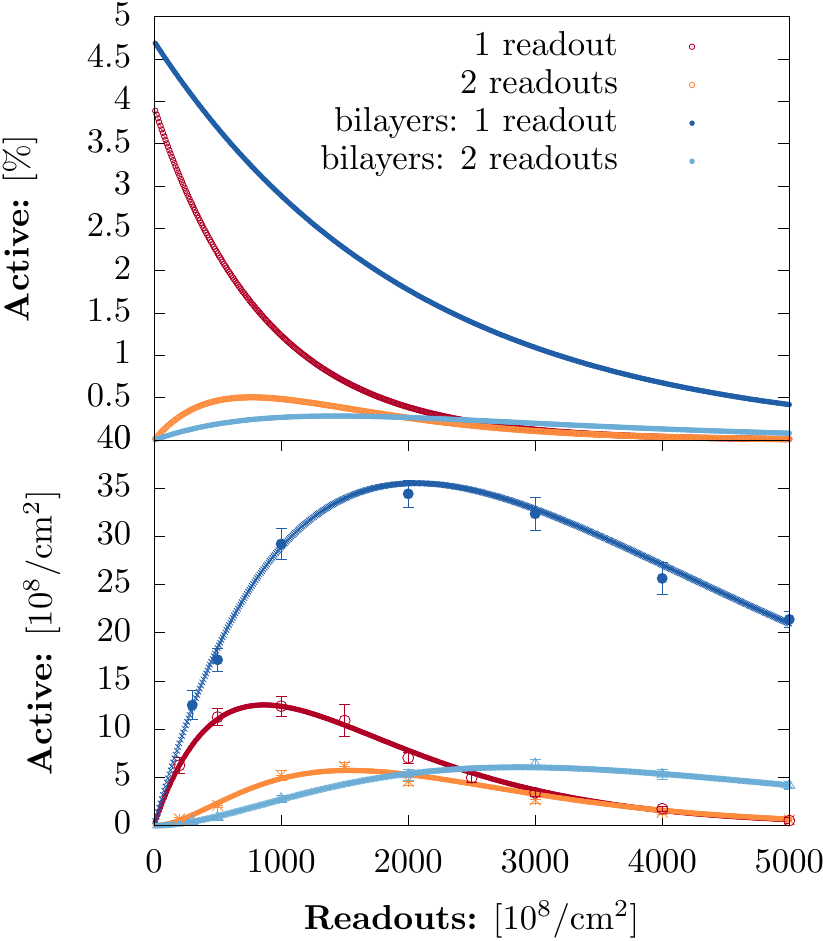}
\caption{\textbf{Dual species doping in 2D - monolayer (reds)/ bilayer (blues)} Active percentage and readout density as a function of the total readout density in 2D doped silicon. Orange and light blue correspond to SFG gates which have two readouts, red and dark blue to excited-ground Heisenberg gates which have one readout. Error bars given by the standard deviation of Monte Carlo simulation results.}
\label{graph_2D}
\end{figure}

\begin{figure}
\includegraphics{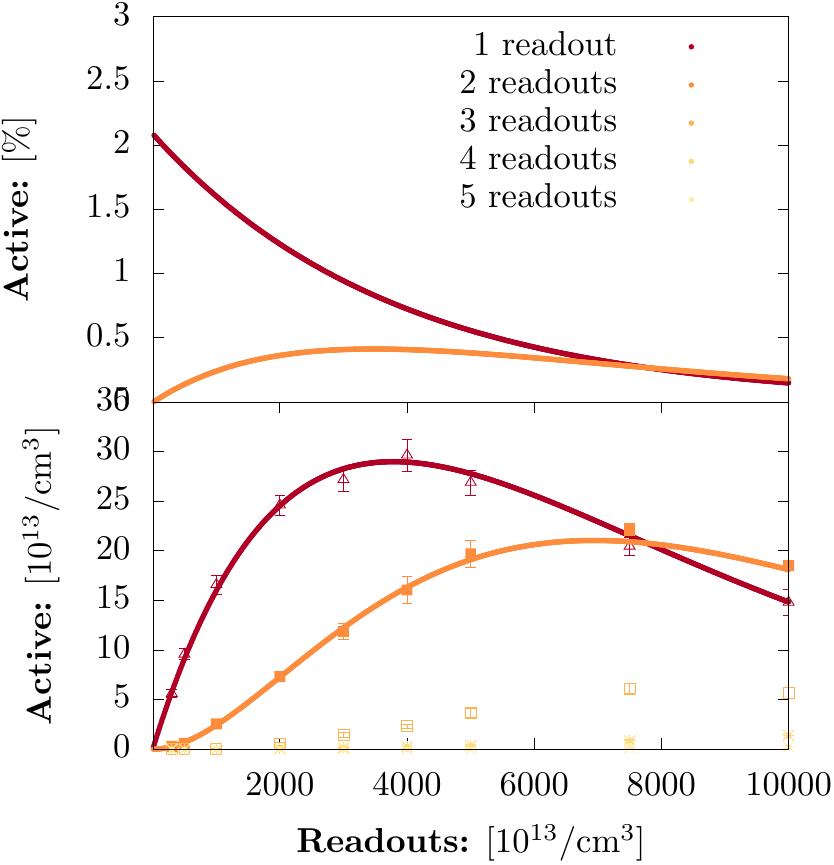}
\caption{\textbf{Dual species doping in 3D} Active percentage and readout density as a function of the total readout density in 3D doped silicon. Red corresponds to Heisenberg gates (1 readout), orange corresponds to SFG gates (2 readouts) and yellows correspond to higher numbers of readouts/qubits. Error bars given by the standard deviation of Monte Carlo simulation results.}
\label{graph_3D}
\end{figure}

The results of Eq.~\ref{eq_Dsfg}, \ref{eq_Dheisex-gd}, \ref{eq_Dheisex-ex} in 2D for the distances corresponding to Si:P and Si:As in Table. \ref{tab_cond} are plotted in Fig.~\ref{graph_2D}. The plotted quantity is the density of readouts which are actively participating in `successful' entangling gate configurations, ie. twice the density of entangling gates in the SFG case and the same density as that of entangling gates for the Heisenberg single readout gate cases. The percentage of active readouts is simply the density of readouts which are a part of an entangling gate divided by the total density of readouts.

In all cases, the total density of controls is maximised by choosing $D_c=\frac{1}{V(R_{cc})}$ from Eq.~\ref{eq_isolation_density}, yielding the viable control density $\frac{1}{e\times\:V(R_{cc})}$ from Eq.~\ref{vdens}, as can be seen on Fig.~\ref{graph_isolation}. The optimal control densities calculated with the distances of Table. \ref{tab_cond} can be seen in Table. \ref{tab_results}.

In 2D, the maximum density of readout dopants (Si:As) which are a part of `successful' SFG gates is 6 $\times 10^8$ dopants per cm\super{2}, which corresponds to a total readout doping density of 1.5 $\times 10^{11}$ dopants per cm\super{2}. The bilayer case shows a negligible increase in the density of readouts which are part of SFG gates, for double the total density (3 $\times 10^{11}$ dopants per cm\super{2}). The percentages of readouts involved in `successful' SFG configurations in both the monolayer and bilayer cases remain below 0.5\%.

The maximum density of readout dopants (Si:As) which are a part of `successful' excited-readout Heisenberg gates is 1.2 $\times$ 10\super{9} dopants per cm\super{2}, corresponding to a total density of 8 $\times$ 10\super{10} dopants per cm\super{2} and 1.5\% active readouts. A very clear increase can be seen in the bilayer case. The maximum density of readout dopants part of `successful' excited-readout Heisenberg gates in bilayers separated by 13.2nm is 3.5 $\times$ 10\super{9} dopants per cm\super{2}, corresponding to 2 $\times$ 10\super{11} dopants per cm\super{2} and 1.8\% active readouts.

The control (Si:P) doping density that achieves maximum density of active readouts is 1.79 $\times$ 10\super{10} dopants per cm\super{2} in the monolayer case and 3.91 $\times$ 10\super{10} dopants per cm\super{2} in the bilayer case. It is possible to achieve up to 3.9\% active readouts in the monolayer case and 4.7\% active readouts in the bilayer case for low doping densities. Thus, using equal total densities of Si:P and Si:As leads to some of the highest active percentages of Si:As contributing to excited-ground Heisenberg gates.

The 3D bulk doped results can be seen in Fig.~\ref{graph_3D}. The maximum density of readout dopants (Si:As) active in excited-ground Heisenberg type gates is 3 $\times$ 10\super{14} dopants per cm\super{3} corresponding to a total doping density of readouts of 3.5 $\times$ 10\super{16} dopants per cm\super{3} and 0.8\% active readouts. Similar to the 2D case, the maximum active readout percentages of 2.1\% can be reached for total readout doping densities comparable to that of the controls (3.2 $\times$ 10\super{15} dopants per cm\super{3}). The maximum density of readout dopants (Si:As) active in SFG type gates is 2 $\times$ 10\super{14} dopants per cm\super{3} corresponding to a total doping density of readouts of 7 $\times$ 10\super{16} dopants per cm\super{3}. It was possible to gain insights about the densities of gates containing three or more readouts from the Monte-Carlo simulation, which may be of interest for different types of quantum gate, e.g. Toffoli gates. Unsurprisingly, they peak at far higher total readout doping densities, such as 7.5 $\times$ 10\super{16} dopants per cm\super{3} for the three readout case (see Fig. \ref{graph_3D}), but provide lower active readout densities, such as 7 $\times$ 10\super{13} dopants per cm\super{3} for the same case. 

Finally, the single species excited-excited Heisenberg gate in 2D yields the highest densities and percentages of active dopants. Active dopant densities of 5.4 $\times$ 10\super{9} dopants per cm\super{2} can be reached for total doping densities of 2.9 $\times$ 10\super{10} dopants per cm\super{2}, correponding to 20\% of dopants being involved in `successful' excited-excited Heisenberg gates! The active percentage maximum is at 27\%, which corresponds to an active density of 3.8 $\times$ 10\super{9} dopants per cm\super{2} and a total doping density of 1.4 $\times$ 10\super{10} dopants per cm\super{2}.

\begin{figure}
\includegraphics{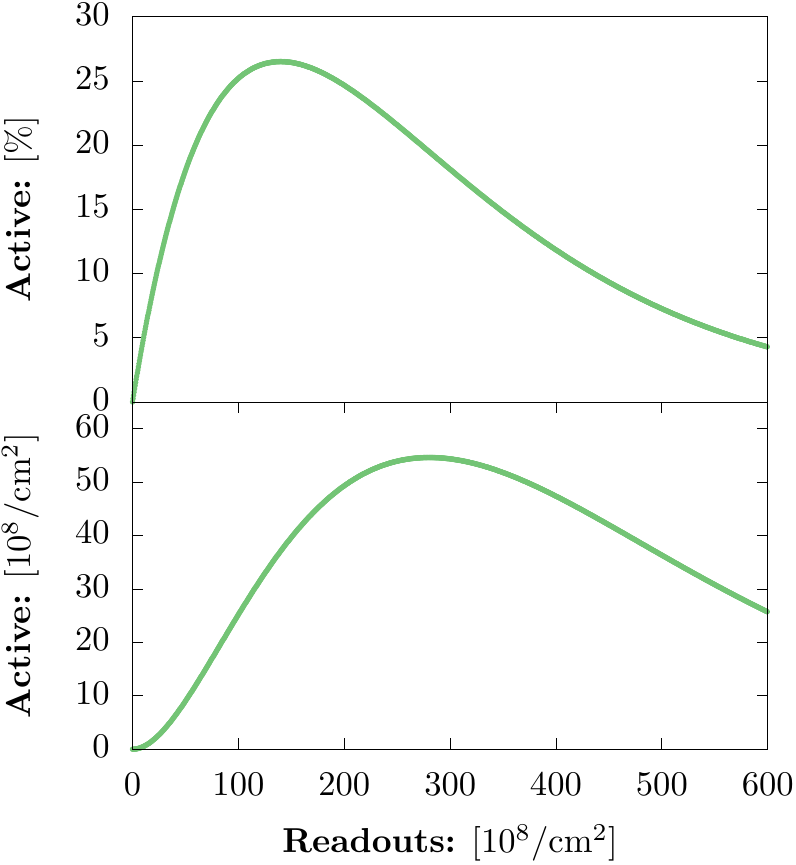}
\caption{\textbf{Single species Heisenberg gate}. The percentage of dopants which are a part of a gate to total number of dopants is far larger than in both other gate types discussed.
\label{graph_dimers}}
\end{figure}

\begin{table}
\centering
\begin{tabular}{c|c|c|c|c|c|}
• & • & $D_{qubits}$ & \% & $D_{As}$ & $D_{P}$ \\ 
\hline 
\\[-1em]
2D & P-As-As & $6 \times 10^{8}$ & $<0.5$ & $1.5 \times 10^{11}$ & $1.8 \times 10^{10}$ \\ 
\hline 
\\[-1em]
• & P-As & $1.2 \times 10^{9}$ & 1.5 & $8 \times 10^{10}$ & $1.8 \times 10^{10}$ \\ 
\hline 
\\[-1em]
• & P-P & $5.4 \times 10^{9}$ & 20 & 0 & $2.9 \times 10^{10}$ \\ 
\hline 
\\[-1em]
3D & SFG & $2 \times 10^{14}$ & $<0.5$ & $7 \times 10^{16}$ & $3.2 \times 10^{15}$ \\ 
\hline 
\\[-1em]
• & P-As & $3 \times 10^{14}$ & 0.8 & $3.5 \times 10^{16}$ & $3.2 \times 10^{15}$ \\ 
\hline
\\[-1em]
bilayer & P-As-As & $6 \times 10^{8}$ & $<0.5$ & $3 \times 10^{11}$ & $3.9 \times 10^{10}$ \\ 
\hline
\\[-1em]
• & P-As & $3.5 \times 10^{9}$ & 1.8 & $2 \times 10^{11}$ & $3.9 \times 10^{10}$ \\ 
\hline
\end{tabular}
\caption{\label{tab_results}Analytical (matching to simulation) results for total densities of control (P) and readout (As) dopants and densities of readouts which are part of gates (D$_{qubits}$). Densities are in dopants per cm$^2$ (in 2D) or cm$^3$ (in 3D).}
\end{table}

\section{Proof of principle experiment: Control of magnetization dynamics by orbital excitation}

As a first step towards the implementation of the optically excited exchange entangling gates we propose a proof of principle experiment to show the control of the magnetization dynamics of the readout (Si:As) donor electrons by the control (Si:P) electron's orbital state, due to enhancement of exchange interactions from orbital excitations. This is the most important building block of the SFG entangling gate. We calculate the quantum many-body magnetization dynamics in the thermodynamic limit using the diagonalization method Moving Average Cluster Expansion (MACE)~\cite{hazzard_many-body_2014}. Because deep donors in silicon act according to the Heisenberg Hamiltonian, this experiment can also be interpreted as a quantum simulation of the two-species S=1/2 Heisenberg anti-ferromagnet\footnote{This has been shown to be an effective low energy description of the half-filled Hubbard model, in the case where the interaction energy between the spins of the lattice is far greater than the hopping strength~\cite{auerbach_interacting_2012,kubo_antiferromagnetic_1975,polatsek_ground-state_1996}.} with quenched disorder.

\subsection{Experimental proposal}

Starting from a spin polarized state we show below that when all donors are in the orbital ground state, the magnetization of Si:As stays constant on observable timescales, while it changes to a vastly different value when the control species is excited to the 2p state. This implies that while exchange interactions are negligible in the orbital ground state, they are large enough to exhibit non-trivial many body dynamics when in the excited state. Realizing this in experiment would at least partially prove the necessary control over two-body interactions needed for the quantum logic gates discussed in the previous sections. 

In experiment, readout of the magnetisation of the Si:As is done via donor-bound exciton spectroscopy (D\sub{0}X spectroscopy)\cite{saeedi_donorbound_2015,morley_review_2014}. This requires electrical detection, which can be achieved in a dilute 2D layer of P and As donors using STM hydrogen lithography to pattern highly conductive metallic-doped phosphorous pads into the same plane and overgrowing a protective thin-film of crystalline silicon with Molecular beam epitaxy. The impurity sheet's metallically doped pads are electrically contacted using electron beam lithography coupled to reactive ion etching to create features which are filled with aluminium using a metal evaporator. The metallic pads are in turn electrically contacted to obtain, in conjunction with terahertz radiation from a Free Electron Laser (FEL) ($\lambda = 31.6 \mu$m for the Si:P 1s to 2p+- transition) and D\sub{0}X spectroscopy, an electrical signal from the 2D Si:As sheet which is a response to the coherent and non-linear excitations of the Si:P electrons. The sample fabrication and electrical detection technique briefly described above and which we have in mind for the experiment we propose here are described in~\cite{crane_fabrication_2018}. This detection technique enables the precision condensed matter samples to remain intact after exposure to a FEL pulse.

The contacted impurity layer in silicon is mounted on the bore of a water-cooled Bitter magnet to Zeeman split the ground state impurity electron spin energies, leading to six pairs of dipole-allowed transitions ($\Delta$m = 0, $\pm$1). Electrical detection of electron spin resonance using D\sub{0}X spectroscopy has been demonstrated for magnetic fields of around 0.35 Tesla~\cite{lo_hybrid_2015}. A donor-bound exciton can be formed by a direct 1.15 eV photon (the silicon indirect bandgap is of 1.17 eV). The photon excites an electron from the valence band, leaving behind a hole. When the electron-hole pair recombine via an Auger recombination process, their energy ejects another electron from the donor site, leaving behind a positively charged donor ion\cite{portis_electron_1953}. To relax all the electron spin states of both donor species to the lower energy spin state, the sample needs to be cooled down in a dilution fridge to milli-Kelvin temperatures. D\sub{0}X optical pumping then initialises all the readouts to the opposite spin state from the controls (and later to read out the occupation of one of the Si:As spin states). Each electron has the initial spin state $\ket{S_{i}}$ with 
\begin{equation}
\ket{S_{i}}=\begin{cases}
\ket{\uparrow}&\mathrm{if}\quad i=\mathrm{readout}\\
\ket{\downarrow}&\mathrm{if}\quad i=\mathrm{control}
\end{cases}
\end{equation} and the whole system is initially in a product state $\ket{\psi_0}$ such that:
\begin{equation}
\ket{\psi_0} =  \ket{S_1\:S_2\:...S_n}.
\end{equation}

When the terahertz frequency FEL is switched on, it illuminates the entire sample and excites the control (Si:P) electrons to the 2p+- state. The experimentally demonstrated decay time of 3D bulk doped Si:P in the 2p+- state in non-isotopic silicon is 200ps~\cite{litvinenko_coherent_2015}, which gives the timescale available to our experiment.

The dopants evolve in time according to the Heisenberg Hamiltonian
\begin{align}
\mathcal{H} = &\frac{1}{2} \sum_{i \neq j} J_{ij}^{AsAs} \textbf{S}^{As}_{i} \textbf{S}^{As}_{j}+\frac{1}{2} \sum_{i \neq j} J_{ij}^{PAs} \textbf{S}^{P}_{i} \textbf{S}^{As}_{j}\notag\\&\quad+\frac{1}{2} \sum_{i \neq j} J_{ij} ^{PP}\textbf{S}^{P}_{i} \textbf{S}^{P}_{j},
\end{align} where we made the species dependence of the interactions explicit. The strength of the interaction $J^{PAs},J^{PP}$ depends explicitly on the orbital state of the control species (Si:P) (see Appendix \ref{app_jex_calc}).
If the densities are high enough, the Si:P wavefunctions will overlap with the Si:As and lead to non-trivial dynamics.

The average magnetization of the readouts (Si:As) is given by
\begin{align}
\braket{S^{z\:\mathrm{As}}(t)} = \braket{\psi_0|\frac{1}{N}\sum_iS^{z\:\mathrm{As}}_i(t)|\psi_0}.
\end{align}
The average spin flip probability is then defined as
\begin{align}
P_{\mathrm{sf}}^{\mathrm{As}}(t) = & \frac{1}{N}\sum_{i} \big\langle\ket{\uparrow}\bra{\uparrow}_i\big\rangle\\
= & \frac{1}{2N} + \braket{S^{z\:\mathrm{As}}(t)},
\end{align} where $\ket{\uparrow}\bra{\uparrow}_i$ is the projector onto the $i$-th spin being in spin up.

To make a differential measurement, we compare the average spin flip probability of the readout dopants when the control dopants are in the ground state with the average spin flip probability when the control dopants are collectively excited to the 2p+- state. As a result, we then expect
\begin{equation}
P_{\mathrm{sf}}^{\mathrm{As}}(t)\rightarrow\begin{cases}
\approx 0 &\mathrm{for}\,\mathrm{control}\,\mathrm{in}\,1s\\
\neq 0 &\mathrm{for}\,\mathrm{control}\,\mathrm{in}\,2p\\
\end{cases},\mathrm{\:for\:late\:times\:t}.
\end{equation}
We can deduce that the non-trivial dynamics of the spin flip probability must have been due to interactions between species as the species themselves are in an eigenstate of their respective Heisenberg Hamiltonian.

In the following, we will now predict the outcome of the experiment using an exact diagonalization technique.

\subsection{Dynamics of the arsenic spin flip probability within MACE}

\begin{figure}
\includegraphics{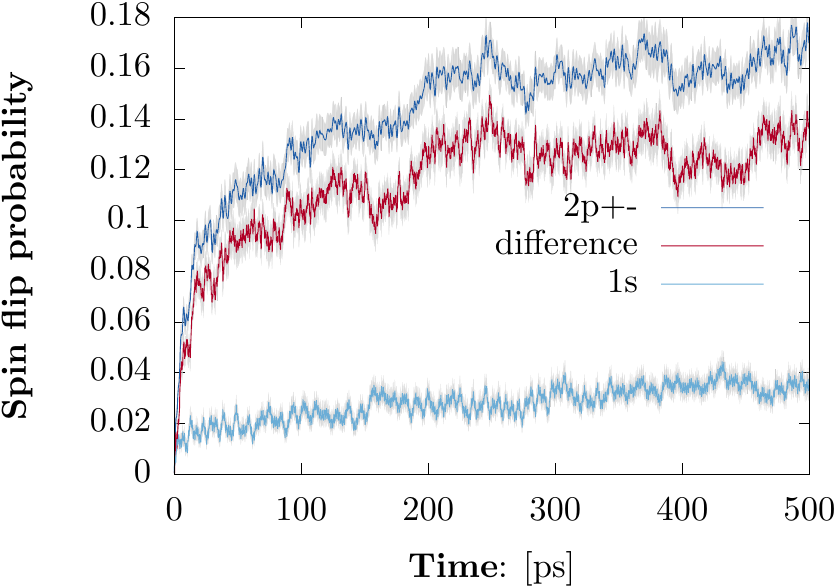}
\caption{\textbf{Average spin flip probability of the Si:As for Si:P and Si:As densities given by $7\times10^{9}$ dopants per cm$^2$ as obtained with MACE, where both species are in the same 2D layer}. Si:As is always in the 1s ground state. Light blue:  P in the 1s ground state, dark blue: P in the 2p+- excited state, red: the absolute value of the difference between both curves. The results are averaged over 400 clusters of size 8 and the shaded areas around the curves are the standard deviation estimated with a jackknife binning analysis of the cluster sampling.}
\label{graph_mace2}
\end{figure}

In order to simulate the experimental outcome, i.e. the magnetization dynamics, we use the Moving Average Cluster Expansion(MACE) technique which has been shown to capture the magnetization dynamics in disordered long-range interacting quantum spin systems realized by cold dipolar molecules~\cite{hazzard_many-body_2014} and Rydberg atoms~\cite{signoles_relaxation_2017}.

MACE assumes that in a system where spins are randomly placed in space,  contributions to the local magnetization dynamics of a particular spin are only made by the spins with which it has the largest interaction exchange energies. Therefore, it is enough to exactly diagonalize this cluster and then average over all such clusters. Convergence is then sought by increasing the cluster size. In our system, convergence was found for cluster sizes 5 and above due to the short-range nature of the interactions. This is in contrast to the large cluster sizes needed in algebraically decaying interactions~\cite{schuckert_nonequilibrium_2018}. Plotted on Fig.~\ref{graph_mace2} are results from cluster size 8 with jackknife error estimates in grey~\cite{ambegaokar_estimating_10}. 

Neglecting the effect of the magnetic field on the strength of the exchange interaction, we calculated that there is no visible change in the magnetisation of the arsenic for fields up to 10 Tesla. We have also checked that there are no significant differences between defining clusters with the largest exchange interactions or the nearest neighbours of the spin of interest.

The difference in the dynamics of the Si:As spin-flip probability between Si:P being in the orbital ground state or the excited state can be seen in Fig.~\ref{graph_mace2} to be 13\% within the 200ps experimentally detected decoherence time of the Si:P 2p+- excited state, for doping densities of both As and P of $7\times10^{9}$ dopants per cm$^2$, which is accessible with current doping techniques. 

This shows that already with only global measurements at hand, non-trivial quantum many-body dynamics could be probed within this dual-doped system. Comparing measurements where Si:P is in the ground and excited state would furthermore show that substantial control over the dynamics of the Si:As spins can be implemented within this scheme, therefore showing that one of the prime requirements of the SFG entangling gate are within reach of current experimental platforms.

\section*{Conclusion}
In this paper we have established the optimum densities of optically controlled entangling gates achievable in randomly doped silicon. In order to determine which spacing between dopant electrons can realistically create entangling gates, we created spatial maps of the Heisenberg interaction exchange energy between same-species and dual species dopants. We focused on Si:P and Si:As because they are, to date, the most well-understood dopants for diffusion onto a silicon surface in ultra-high vacuum. 

Respecting these conditions we obtained matching results in both the nearest neighbour treatment of a Poisson point process and a Monte-Carlo simulation. Densities of entangling gates in 2D were in the low $10^9$ dopants per cm\super{2} and in 3D were in the low $10^{13}$ dopants per cm\super{3}, corresponding to relatively low percentages of active dopants. However, by considering same-species `Heisenberg' gates, similar densities corresponded to percentages up to 27\% of dopants involved in `successful' gate structures. We also showed that in 3D there arise situations in which gates with three and four qubits reach significant densities.  Finally that by dividing the donors into two parallel 2D layers the densities of gates increase still further.

Finally, we proposed a proof of principle experiment aimed at demonstrating the onset of two-body (entangling) interactions caused by the optical excitation of one of the species in a 2D randomly doped sample. The magnetisation dynamics of the Si:As donor electron spins were calculated using the Moving Average Cluster Expansion technique, depending on the orbital state of the Si:P donor electrons. For low but equal densities of Si:P and Si:As, we found a large difference in the time evolution of the spin flip probability of more than 10\% between the cases where phosphorous was in the ground or excited state.

This work can be extended in various ways. Single donor placement techniques such as hydrogen lithography could be used to create two or three dopant structures in silicon to verify the maps of the exchange interaction energy calculated here.  The theory we have developed can be applied to different dopant species such as selenium or acceptors such as boron, could be modified to accommodate hard-sphere configurations and which would be relevant for Rydberg atoms which deviate from pure Poissonian statistics in the blockaded regime~\cite{RevModPhys.82.2313}.

\begin{acknowledgments}
We thankfully acknowledge discussions with K.~Stockbridge, B.~N.~Murdin, N.~J.~Curson, G.~Matmon and R.~Crane. A.S. thanks the London Centre for Nanotechnology for hospitality. We gratefully acknowledge the financial support from the UK Engineering and Physical Sciences Research Council [COMPASSS/ADDRFSS, Grant No. EP/M009564/1]. A.S. acknowledges financial support from the International Max Planck Research School for Quantum Science and Technology (IMPRS-QST) funded by the Max Planck Gesellschaft(MPG).
 \end{acknowledgments}
\bibliography{bibliography}

\begin{thebibliography}{77}%
\makeatletter
\providecommand \@ifxundefined [1]{%
 \@ifx{#1\undefined}
}%
\providecommand \@ifnum [1]{%
 \ifnum #1\expandafter \@firstoftwo
 \else \expandafter \@secondoftwo
 \fi
}%
\providecommand \@ifx [1]{%
 \ifx #1\expandafter \@firstoftwo
 \else \expandafter \@secondoftwo
 \fi
}%
\providecommand \natexlab [1]{#1}%
\providecommand \enquote  [1]{``#1''}%
\providecommand \bibnamefont  [1]{#1}%
\providecommand \bibfnamefont [1]{#1}%
\providecommand \citenamefont [1]{#1}%
\providecommand \href@noop [0]{\@secondoftwo}%
\providecommand \href [0]{\begingroup \@sanitize@url \@href}%
\providecommand \@href[1]{\@@startlink{#1}\@@href}%
\providecommand \@@href[1]{\endgroup#1\@@endlink}%
\providecommand \@sanitize@url [0]{\catcode `\\12\catcode `\$12\catcode
  `\&12\catcode `\#12\catcode `\^12\catcode `\_12\catcode `\%12\relax}%
\providecommand \@@startlink[1]{}%
\providecommand \@@endlink[0]{}%
\providecommand \url  [0]{\begingroup\@sanitize@url \@url }%
\providecommand \@url [1]{\endgroup\@href {#1}{\urlprefix }}%
\providecommand \urlprefix  [0]{URL }%
\providecommand \Eprint [0]{\href }%
\providecommand \doibase [0]{http://dx.doi.org/}%
\providecommand \selectlanguage [0]{\@gobble}%
\providecommand \bibinfo  [0]{\@secondoftwo}%
\providecommand \bibfield  [0]{\@secondoftwo}%
\providecommand \translation [1]{[#1]}%
\providecommand \BibitemOpen [0]{}%
\providecommand \bibitemStop [0]{}%
\providecommand \bibitemNoStop [0]{.\EOS\space}%
\providecommand \EOS [0]{\spacefactor3000\relax}%
\providecommand \BibitemShut  [1]{\csname bibitem#1\endcsname}%
\let\auto@bib@innerbib\@empty
\bibitem [{\citenamefont {Shor}(1994)}]{shor_algorithms_1994}%
  \BibitemOpen
  \bibfield  {author} {\bibinfo {author} {\bibfnamefont {P.~W.}\ \bibnamefont
  {Shor}},\ }in\ \href {\doibase 10.1109/SFCS.1994.365700} {\emph {\bibinfo
  {booktitle} {Proceedings 35th Annual Symposium on Foundations of Computer
  Science}}}\ (\bibinfo {year} {1994})\ pp.\ \bibinfo {pages}
  {124--134}\BibitemShut {NoStop}%
\bibitem [{\citenamefont {Grover}(1996)}]{grover_fast_1996}%
  \BibitemOpen
  \bibfield  {author} {\bibinfo {author} {\bibfnamefont {L.~K.}\ \bibnamefont
  {Grover}},\ }\href {\doibase 10.1145/237814.237866} {\bibfield  {journal}
  {\bibinfo  {journal} {Proceedings of the Twenty-eighth Annual ACM Symposium
  on Theory of Computing}\ }\bibinfo {series} {STOC '96},\ \bibinfo {pages}
  {212} (\bibinfo {year} {1996})}\BibitemShut {NoStop}%
\bibitem [{\citenamefont {Biamonte}\ \emph {et~al.}(2017)\citenamefont
  {Biamonte}, \citenamefont {Wittek}, \citenamefont {Pancotti}, \citenamefont
  {Rebentrost}, \citenamefont {Wiebe},\ and\ \citenamefont
  {Lloyd}}]{biamonte_quantum_2017}%
  \BibitemOpen
  \bibfield  {author} {\bibinfo {author} {\bibfnamefont {J.}~\bibnamefont
  {Biamonte}}, \bibinfo {author} {\bibfnamefont {P.}~\bibnamefont {Wittek}},
  \bibinfo {author} {\bibfnamefont {N.}~\bibnamefont {Pancotti}}, \bibinfo
  {author} {\bibfnamefont {P.}~\bibnamefont {Rebentrost}}, \bibinfo {author}
  {\bibfnamefont {N.}~\bibnamefont {Wiebe}}, \ and\ \bibinfo {author}
  {\bibfnamefont {S.}~\bibnamefont {Lloyd}},\ }\href {\doibase
  10.1038/nature23474} {\bibfield  {journal} {\bibinfo  {journal} {Nature}\
  }\textbf {\bibinfo {volume} {549}},\ \bibinfo {pages} {195} (\bibinfo {year}
  {2017})}\BibitemShut {NoStop}%
\bibitem [{\citenamefont {Feynman}(1982)}]{feynman_simulating_1982}%
  \BibitemOpen
  \bibfield  {author} {\bibinfo {author} {\bibfnamefont {R.~P.}\ \bibnamefont
  {Feynman}},\ }\href {\doibase 10.1007/BF02650179} {\bibfield  {journal}
  {\bibinfo  {journal} {International Journal of Theoretical Physics}\ }\textbf
  {\bibinfo {volume} {21}},\ \bibinfo {pages} {467} (\bibinfo {year}
  {1982})}\BibitemShut {NoStop}%
\bibitem [{\citenamefont {King}\ \emph {et~al.}(2018)\citenamefont {King},
  \citenamefont {Carrasquilla}, \citenamefont {Raymond}, \citenamefont
  {Ozfidan}, \citenamefont {Andriyash}, \citenamefont {Berkley}, \citenamefont
  {Reis}, \citenamefont {Lanting}, \citenamefont {Harris}, \citenamefont
  {Altomare}, \citenamefont {Boothby}, \citenamefont {Bunyk}, \citenamefont
  {Enderud}, \citenamefont {Fréchette}, \citenamefont {Hoskinson},
  \citenamefont {Ladizinsky}, \citenamefont {Oh}, \citenamefont
  {Poulin-Lamarre}, \citenamefont {Rich}, \citenamefont {Sato}, \citenamefont
  {Smirnov}, \citenamefont {Swenson}, \citenamefont {Volkmann}, \citenamefont
  {Whittaker}, \citenamefont {Yao}, \citenamefont {Ladizinsky}, \citenamefont
  {Johnson}, \citenamefont {Hilton},\ and\ \citenamefont
  {Amin}}]{king_observation_2018}%
  \BibitemOpen
  \bibfield  {author} {\bibinfo {author} {\bibfnamefont {A.~D.}\ \bibnamefont
  {King}}, \bibinfo {author} {\bibfnamefont {J.}~\bibnamefont {Carrasquilla}},
  \bibinfo {author} {\bibfnamefont {J.}~\bibnamefont {Raymond}}, \bibinfo
  {author} {\bibfnamefont {I.}~\bibnamefont {Ozfidan}}, \bibinfo {author}
  {\bibfnamefont {E.}~\bibnamefont {Andriyash}}, \bibinfo {author}
  {\bibfnamefont {A.}~\bibnamefont {Berkley}}, \bibinfo {author} {\bibfnamefont
  {M.}~\bibnamefont {Reis}}, \bibinfo {author} {\bibfnamefont {T.}~\bibnamefont
  {Lanting}}, \bibinfo {author} {\bibfnamefont {R.}~\bibnamefont {Harris}},
  \bibinfo {author} {\bibfnamefont {F.}~\bibnamefont {Altomare}}, \bibinfo
  {author} {\bibfnamefont {K.}~\bibnamefont {Boothby}}, \bibinfo {author}
  {\bibfnamefont {P.~I.}\ \bibnamefont {Bunyk}}, \bibinfo {author}
  {\bibfnamefont {C.}~\bibnamefont {Enderud}}, \bibinfo {author} {\bibfnamefont
  {A.}~\bibnamefont {Fréchette}}, \bibinfo {author} {\bibfnamefont
  {E.}~\bibnamefont {Hoskinson}}, \bibinfo {author} {\bibfnamefont
  {N.}~\bibnamefont {Ladizinsky}}, \bibinfo {author} {\bibfnamefont
  {T.}~\bibnamefont {Oh}}, \bibinfo {author} {\bibfnamefont {G.}~\bibnamefont
  {Poulin-Lamarre}}, \bibinfo {author} {\bibfnamefont {C.}~\bibnamefont
  {Rich}}, \bibinfo {author} {\bibfnamefont {Y.}~\bibnamefont {Sato}}, \bibinfo
  {author} {\bibfnamefont {A.~Y.}\ \bibnamefont {Smirnov}}, \bibinfo {author}
  {\bibfnamefont {L.~J.}\ \bibnamefont {Swenson}}, \bibinfo {author}
  {\bibfnamefont {M.~H.}\ \bibnamefont {Volkmann}}, \bibinfo {author}
  {\bibfnamefont {J.}~\bibnamefont {Whittaker}}, \bibinfo {author}
  {\bibfnamefont {J.}~\bibnamefont {Yao}}, \bibinfo {author} {\bibfnamefont
  {E.}~\bibnamefont {Ladizinsky}}, \bibinfo {author} {\bibfnamefont {M.~W.}\
  \bibnamefont {Johnson}}, \bibinfo {author} {\bibfnamefont {J.}~\bibnamefont
  {Hilton}}, \ and\ \bibinfo {author} {\bibfnamefont {M.~H.}\ \bibnamefont
  {Amin}},\ }\href {\doibase 10.1038/s41586-018-0410-x} {\bibfield  {journal}
  {\bibinfo  {journal} {Nature}\ }\textbf {\bibinfo {volume} {560}},\ \bibinfo
  {pages} {456} (\bibinfo {year} {2018})}\BibitemShut {NoStop}%
\bibitem [{\citenamefont {Braumüller}\ \emph {et~al.}(2017)\citenamefont
  {Braumüller}, \citenamefont {Marthaler}, \citenamefont {Schneider},
  \citenamefont {Stehli}, \citenamefont {Rotzinger}, \citenamefont {Weides},\
  and\ \citenamefont {Ustinov}}]{braumuller_analog_2017}%
  \BibitemOpen
  \bibfield  {author} {\bibinfo {author} {\bibfnamefont {J.}~\bibnamefont
  {Braumüller}}, \bibinfo {author} {\bibfnamefont {M.}~\bibnamefont
  {Marthaler}}, \bibinfo {author} {\bibfnamefont {A.}~\bibnamefont
  {Schneider}}, \bibinfo {author} {\bibfnamefont {A.}~\bibnamefont {Stehli}},
  \bibinfo {author} {\bibfnamefont {H.}~\bibnamefont {Rotzinger}}, \bibinfo
  {author} {\bibfnamefont {M.}~\bibnamefont {Weides}}, \ and\ \bibinfo {author}
  {\bibfnamefont {A.~V.}\ \bibnamefont {Ustinov}},\ }\href {\doibase
  10.1038/s41467-017-00894-w} {\bibfield  {journal} {\bibinfo  {journal}
  {Nature Communications}\ }\textbf {\bibinfo {volume} {8}},\ \bibinfo {pages}
  {779} (\bibinfo {year} {2017})}\BibitemShut {NoStop}%
\bibitem [{\citenamefont {Lamata}\ \emph {et~al.}(2018)\citenamefont {Lamata},
  \citenamefont {Parra-Rodriguez}, \citenamefont {Sanz},\ and\ \citenamefont
  {Solano}}]{lamata_analog_2018}%
  \BibitemOpen
  \bibfield  {author} {\bibinfo {author} {\bibfnamefont {L.}~\bibnamefont
  {Lamata}}, \bibinfo {author} {\bibfnamefont {A.}~\bibnamefont
  {Parra-Rodriguez}}, \bibinfo {author} {\bibfnamefont {M.}~\bibnamefont
  {Sanz}}, \ and\ \bibinfo {author} {\bibfnamefont {E.}~\bibnamefont
  {Solano}},\ }\href {\doibase 10.1080/23746149.2018.1457981} {\bibfield
  {journal} {\bibinfo  {journal} {Advances in Physics: X}\ }\textbf {\bibinfo
  {volume} {3}},\ \bibinfo {pages} {1457981} (\bibinfo {year}
  {2018})}\BibitemShut {NoStop}%
\bibitem [{\citenamefont {H\"affner}\ \emph {et~al.}(2008)\citenamefont
  {H\"affner}, \citenamefont {Roos},\ and\ \citenamefont
  {Blatt}}]{haffner_quantum_2008}%
  \BibitemOpen
  \bibfield  {author} {\bibinfo {author} {\bibfnamefont {H.}~\bibnamefont
  {H\"affner}}, \bibinfo {author} {\bibfnamefont {C.~F.}\ \bibnamefont {Roos}},
  \ and\ \bibinfo {author} {\bibfnamefont {R.}~\bibnamefont {Blatt}},\ }\href
  {\doibase 10.1016/j.physrep.2008.09.003} {\bibfield  {journal} {\bibinfo
  {journal} {Physics Reports}\ }\textbf {\bibinfo {volume} {469}},\ \bibinfo
  {pages} {155} (\bibinfo {year} {2008})}\BibitemShut {NoStop}%
\bibitem [{\citenamefont {Zhang}\ \emph {et~al.}(2017)\citenamefont {Zhang},
  \citenamefont {Pagano}, \citenamefont {Hess}, \citenamefont {Kyprianidis},
  \citenamefont {Becker}, \citenamefont {Kaplan}, \citenamefont {Gorshkov},
  \citenamefont {Gong},\ and\ \citenamefont {Monroe}}]{zhang_observation_2017}%
  \BibitemOpen
  \bibfield  {author} {\bibinfo {author} {\bibfnamefont {J.}~\bibnamefont
  {Zhang}}, \bibinfo {author} {\bibfnamefont {G.}~\bibnamefont {Pagano}},
  \bibinfo {author} {\bibfnamefont {P.~W.}\ \bibnamefont {Hess}}, \bibinfo
  {author} {\bibfnamefont {A.}~\bibnamefont {Kyprianidis}}, \bibinfo {author}
  {\bibfnamefont {P.}~\bibnamefont {Becker}}, \bibinfo {author} {\bibfnamefont
  {H.}~\bibnamefont {Kaplan}}, \bibinfo {author} {\bibfnamefont {A.~V.}\
  \bibnamefont {Gorshkov}}, \bibinfo {author} {\bibfnamefont {Z.-X.}\
  \bibnamefont {Gong}}, \ and\ \bibinfo {author} {\bibfnamefont
  {C.}~\bibnamefont {Monroe}},\ }\href {\doibase 10.1038/nature24654}
  {\bibfield  {journal} {\bibinfo  {journal} {Nature}\ }\textbf {\bibinfo
  {volume} {551}},\ \bibinfo {pages} {601} (\bibinfo {year}
  {2017})}\BibitemShut {NoStop}%
\bibitem [{\citenamefont {Greiner}\ \emph {et~al.}(2002)\citenamefont
  {Greiner}, \citenamefont {Mandel}, \citenamefont {Esslinger}, \citenamefont
  {Hansch},\ and\ \citenamefont {Bloch}}]{greiner_quantum_2002}%
  \BibitemOpen
  \bibfield  {author} {\bibinfo {author} {\bibfnamefont {M.}~\bibnamefont
  {Greiner}}, \bibinfo {author} {\bibfnamefont {O.}~\bibnamefont {Mandel}},
  \bibinfo {author} {\bibfnamefont {T.}~\bibnamefont {Esslinger}}, \bibinfo
  {author} {\bibfnamefont {T.~W.}\ \bibnamefont {Hansch}}, \ and\ \bibinfo
  {author} {\bibfnamefont {I.}~\bibnamefont {Bloch}},\ }\href {\doibase
  10.1038/415039a} {\bibfield  {journal} {\bibinfo  {journal} {Nature}\
  }\textbf {\bibinfo {volume} {415}},\ \bibinfo {pages} {39} (\bibinfo {year}
  {2002})}\BibitemShut {NoStop}%
\bibitem [{\citenamefont {Gross}\ and\ \citenamefont
  {Bloch}(2017)}]{gross_quantum_2017}%
  \BibitemOpen
  \bibfield  {author} {\bibinfo {author} {\bibfnamefont {C.}~\bibnamefont
  {Gross}}\ and\ \bibinfo {author} {\bibfnamefont {I.}~\bibnamefont {Bloch}},\
  }\href {\doibase 10.1126/science.aal3837} {\bibfield  {journal} {\bibinfo
  {journal} {Science}\ }\textbf {\bibinfo {volume} {357}},\ \bibinfo {pages}
  {995} (\bibinfo {year} {2017})}\BibitemShut {NoStop}%
\bibitem [{\citenamefont {Mazurenko}\ \emph {et~al.}(2017)\citenamefont
  {Mazurenko}, \citenamefont {Chiu}, \citenamefont {Ji}, \citenamefont
  {Parsons}, \citenamefont {Kanász-Nagy}, \citenamefont {Schmidt},
  \citenamefont {Grusdt}, \citenamefont {Demler}, \citenamefont {Greif},\ and\
  \citenamefont {Greiner}}]{mazurenko_cold-atom_2017}%
  \BibitemOpen
  \bibfield  {author} {\bibinfo {author} {\bibfnamefont {A.}~\bibnamefont
  {Mazurenko}}, \bibinfo {author} {\bibfnamefont {C.~S.}\ \bibnamefont {Chiu}},
  \bibinfo {author} {\bibfnamefont {G.}~\bibnamefont {Ji}}, \bibinfo {author}
  {\bibfnamefont {M.~F.}\ \bibnamefont {Parsons}}, \bibinfo {author}
  {\bibfnamefont {M.}~\bibnamefont {Kanász-Nagy}}, \bibinfo {author}
  {\bibfnamefont {R.}~\bibnamefont {Schmidt}}, \bibinfo {author} {\bibfnamefont
  {F.}~\bibnamefont {Grusdt}}, \bibinfo {author} {\bibfnamefont
  {E.}~\bibnamefont {Demler}}, \bibinfo {author} {\bibfnamefont
  {D.}~\bibnamefont {Greif}}, \ and\ \bibinfo {author} {\bibfnamefont
  {M.}~\bibnamefont {Greiner}},\ }\href {\doibase 10.1038/nature22362}
  {\bibfield  {journal} {\bibinfo  {journal} {Nature}\ }\textbf {\bibinfo
  {volume} {545}},\ \bibinfo {pages} {462} (\bibinfo {year}
  {2017})}\BibitemShut {NoStop}%
\bibitem [{\citenamefont {Bernien}\ \emph {et~al.}(2017)\citenamefont
  {Bernien}, \citenamefont {Schwartz}, \citenamefont {Keesling}, \citenamefont
  {Levine}, \citenamefont {Omran}, \citenamefont {Pichler}, \citenamefont
  {Choi}, \citenamefont {Zibrov}, \citenamefont {Endres}, \citenamefont
  {Greiner}, \citenamefont {Vuletić},\ and\ \citenamefont
  {Lukin}}]{bernien_probing_2017}%
  \BibitemOpen
  \bibfield  {author} {\bibinfo {author} {\bibfnamefont {H.}~\bibnamefont
  {Bernien}}, \bibinfo {author} {\bibfnamefont {S.}~\bibnamefont {Schwartz}},
  \bibinfo {author} {\bibfnamefont {A.}~\bibnamefont {Keesling}}, \bibinfo
  {author} {\bibfnamefont {H.}~\bibnamefont {Levine}}, \bibinfo {author}
  {\bibfnamefont {A.}~\bibnamefont {Omran}}, \bibinfo {author} {\bibfnamefont
  {H.}~\bibnamefont {Pichler}}, \bibinfo {author} {\bibfnamefont
  {S.}~\bibnamefont {Choi}}, \bibinfo {author} {\bibfnamefont {A.~S.}\
  \bibnamefont {Zibrov}}, \bibinfo {author} {\bibfnamefont {M.}~\bibnamefont
  {Endres}}, \bibinfo {author} {\bibfnamefont {M.}~\bibnamefont {Greiner}},
  \bibinfo {author} {\bibfnamefont {V.}~\bibnamefont {Vuletić}}, \ and\
  \bibinfo {author} {\bibfnamefont {M.~D.}\ \bibnamefont {Lukin}},\ }\href
  {\doibase 10.1038/nature24622} {\bibfield  {journal} {\bibinfo  {journal}
  {Nature}\ }\textbf {\bibinfo {volume} {551}},\ \bibinfo {pages} {579}
  (\bibinfo {year} {2017})}\BibitemShut {NoStop}%
\bibitem [{\citenamefont {Sau}(2017)}]{sau_viewpoint_2017}%
  \BibitemOpen
  \bibfield  {author} {\bibinfo {author} {\bibfnamefont {J.}~\bibnamefont
  {Sau}},\ }\href {https://physics.aps.org/articles/v10/68} {\bibfield
  {journal} {\bibinfo  {journal} {Physics}\ }\textbf {\bibinfo {volume} {10}},\
  \bibinfo {pages} {68} (\bibinfo {year} {2017})}\BibitemShut {NoStop}%
\bibitem [{\citenamefont {Lahtinen}\ and\ \citenamefont
  {Pachos}(2017)}]{lahtinen_short_2017}%
  \BibitemOpen
  \bibfield  {author} {\bibinfo {author} {\bibfnamefont {V.}~\bibnamefont
  {Lahtinen}}\ and\ \bibinfo {author} {\bibfnamefont {J.~K.}\ \bibnamefont
  {Pachos}},\ }\href {\doibase 10.21468/SciPostPhys.3.3.021} {\bibfield
  {journal} {\bibinfo  {journal} {SciPost Phys.}\ }\textbf {\bibinfo {volume}
  {3}},\ \bibinfo {pages} {021} (\bibinfo {year} {2017})}\BibitemShut {NoStop}%
\bibitem [{\citenamefont {Devoret}\ \emph {et~al.}(2004)\citenamefont
  {Devoret}, \citenamefont {Wallraff},\ and\ \citenamefont
  {Martinis}}]{devoret_superconducting_2004}%
  \BibitemOpen
  \bibfield  {author} {\bibinfo {author} {\bibfnamefont {M.~H.}\ \bibnamefont
  {Devoret}}, \bibinfo {author} {\bibfnamefont {A.}~\bibnamefont {Wallraff}}, \
  and\ \bibinfo {author} {\bibfnamefont {J.~M.}\ \bibnamefont {Martinis}},\
  }\href@noop {} {\  (\bibinfo {year} {2004})},\ \Eprint
  {http://arxiv.org/abs/arXiv:cond-mat/0411174} {arXiv:cond-mat/0411174}
  \BibitemShut {NoStop}%
\bibitem [{\citenamefont {Barends}\ \emph {et~al.}(2014)\citenamefont
  {Barends}, \citenamefont {Kelly}, \citenamefont {Megrant}, \citenamefont
  {Veitia}, \citenamefont {Sank}, \citenamefont {Jeffrey}, \citenamefont
  {White}, \citenamefont {Mutus}, \citenamefont {Fowler}, \citenamefont
  {Campbell}, \citenamefont {Chen}, \citenamefont {Chen}, \citenamefont
  {Chiaro}, \citenamefont {Dunsworth}, \citenamefont {Neill}, \citenamefont
  {O’Malley}, \citenamefont {Roushan}, \citenamefont {Vainsencher},
  \citenamefont {Wenner}, \citenamefont {Korotkov}, \citenamefont {Cleland},\
  and\ \citenamefont {Martinis}}]{barends_superconducting_2014}%
  \BibitemOpen
  \bibfield  {author} {\bibinfo {author} {\bibfnamefont {R.}~\bibnamefont
  {Barends}}, \bibinfo {author} {\bibfnamefont {J.}~\bibnamefont {Kelly}},
  \bibinfo {author} {\bibfnamefont {A.}~\bibnamefont {Megrant}}, \bibinfo
  {author} {\bibfnamefont {A.}~\bibnamefont {Veitia}}, \bibinfo {author}
  {\bibfnamefont {D.}~\bibnamefont {Sank}}, \bibinfo {author} {\bibfnamefont
  {E.}~\bibnamefont {Jeffrey}}, \bibinfo {author} {\bibfnamefont {T.~C.}\
  \bibnamefont {White}}, \bibinfo {author} {\bibfnamefont {J.}~\bibnamefont
  {Mutus}}, \bibinfo {author} {\bibfnamefont {A.~G.}\ \bibnamefont {Fowler}},
  \bibinfo {author} {\bibfnamefont {B.}~\bibnamefont {Campbell}}, \bibinfo
  {author} {\bibfnamefont {Y.}~\bibnamefont {Chen}}, \bibinfo {author}
  {\bibfnamefont {Z.}~\bibnamefont {Chen}}, \bibinfo {author} {\bibfnamefont
  {B.}~\bibnamefont {Chiaro}}, \bibinfo {author} {\bibfnamefont
  {A.}~\bibnamefont {Dunsworth}}, \bibinfo {author} {\bibfnamefont
  {C.}~\bibnamefont {Neill}}, \bibinfo {author} {\bibfnamefont
  {P.}~\bibnamefont {O’Malley}}, \bibinfo {author} {\bibfnamefont
  {P.}~\bibnamefont {Roushan}}, \bibinfo {author} {\bibfnamefont
  {A.}~\bibnamefont {Vainsencher}}, \bibinfo {author} {\bibfnamefont
  {J.}~\bibnamefont {Wenner}}, \bibinfo {author} {\bibfnamefont {A.~N.}\
  \bibnamefont {Korotkov}}, \bibinfo {author} {\bibfnamefont {A.~N.}\
  \bibnamefont {Cleland}}, \ and\ \bibinfo {author} {\bibfnamefont {J.~M.}\
  \bibnamefont {Martinis}},\ }\href {\doibase 10.1038/nature13171} {\bibfield
  {journal} {\bibinfo  {journal} {Nature}\ }\textbf {\bibinfo {volume} {508}},\
  \bibinfo {pages} {500} (\bibinfo {year} {2014})}\BibitemShut {NoStop}%
\bibitem [{\citenamefont {Kelly}\ \emph {et~al.}(2015)\citenamefont {Kelly},
  \citenamefont {Barends}, \citenamefont {Fowler}, \citenamefont {Megrant},
  \citenamefont {Jeffrey}, \citenamefont {White}, \citenamefont {Sank},
  \citenamefont {Mutus}, \citenamefont {Campbell}, \citenamefont {Chen},
  \citenamefont {Chen}, \citenamefont {Chiaro}, \citenamefont {Dunsworth},
  \citenamefont {Hoi}, \citenamefont {Neill}, \citenamefont {O’Malley},
  \citenamefont {Quintana}, \citenamefont {Roushan}, \citenamefont
  {Vainsencher}, \citenamefont {Wenner}, \citenamefont {Cleland},\ and\
  \citenamefont {Martinis}}]{kelly_state_2015}%
  \BibitemOpen
  \bibfield  {author} {\bibinfo {author} {\bibfnamefont {J.}~\bibnamefont
  {Kelly}}, \bibinfo {author} {\bibfnamefont {R.}~\bibnamefont {Barends}},
  \bibinfo {author} {\bibfnamefont {A.~G.}\ \bibnamefont {Fowler}}, \bibinfo
  {author} {\bibfnamefont {A.}~\bibnamefont {Megrant}}, \bibinfo {author}
  {\bibfnamefont {E.}~\bibnamefont {Jeffrey}}, \bibinfo {author} {\bibfnamefont
  {T.~C.}\ \bibnamefont {White}}, \bibinfo {author} {\bibfnamefont
  {D.}~\bibnamefont {Sank}}, \bibinfo {author} {\bibfnamefont {J.~Y.}\
  \bibnamefont {Mutus}}, \bibinfo {author} {\bibfnamefont {B.}~\bibnamefont
  {Campbell}}, \bibinfo {author} {\bibfnamefont {Y.}~\bibnamefont {Chen}},
  \bibinfo {author} {\bibfnamefont {Z.}~\bibnamefont {Chen}}, \bibinfo {author}
  {\bibfnamefont {B.}~\bibnamefont {Chiaro}}, \bibinfo {author} {\bibfnamefont
  {A.}~\bibnamefont {Dunsworth}}, \bibinfo {author} {\bibfnamefont {I.-C.}\
  \bibnamefont {Hoi}}, \bibinfo {author} {\bibfnamefont {C.}~\bibnamefont
  {Neill}}, \bibinfo {author} {\bibfnamefont {P.~J.~J.}\ \bibnamefont
  {O’Malley}}, \bibinfo {author} {\bibfnamefont {C.}~\bibnamefont
  {Quintana}}, \bibinfo {author} {\bibfnamefont {P.}~\bibnamefont {Roushan}},
  \bibinfo {author} {\bibfnamefont {A.}~\bibnamefont {Vainsencher}}, \bibinfo
  {author} {\bibfnamefont {J.}~\bibnamefont {Wenner}}, \bibinfo {author}
  {\bibfnamefont {A.~N.}\ \bibnamefont {Cleland}}, \ and\ \bibinfo {author}
  {\bibfnamefont {J.~M.}\ \bibnamefont {Martinis}},\ }\href {\doibase
  10.1038/nature14270} {\bibfield  {journal} {\bibinfo  {journal} {Nature}\
  }\textbf {\bibinfo {volume} {519}},\ \bibinfo {pages} {66} (\bibinfo {year}
  {2015})}\BibitemShut {NoStop}%
\bibitem [{\citenamefont {Childress}\ and\ \citenamefont
  {Hanson}(2013)}]{childress_diamond_2013}%
  \BibitemOpen
  \bibfield  {author} {\bibinfo {author} {\bibfnamefont {L.}~\bibnamefont
  {Childress}}\ and\ \bibinfo {author} {\bibfnamefont {R.}~\bibnamefont
  {Hanson}},\ }\href {\doibase 10.1557/mrs.2013.20} {\bibfield  {journal}
  {\bibinfo  {journal} {MRS Bulletin}\ }\textbf {\bibinfo {volume} {38}},\
  \bibinfo {pages} {134} (\bibinfo {year} {2013})}\BibitemShut {NoStop}%
\bibitem [{\citenamefont {Mi}\ \emph {et~al.}(2017)\citenamefont {Mi},
  \citenamefont {Cady}, \citenamefont {Zajac}, \citenamefont {Stehlik},
  \citenamefont {Edge},\ and\ \citenamefont {Petta}}]{mi_circuit_2017}%
  \BibitemOpen
  \bibfield  {author} {\bibinfo {author} {\bibfnamefont {X.}~\bibnamefont
  {Mi}}, \bibinfo {author} {\bibfnamefont {J.~V.}\ \bibnamefont {Cady}},
  \bibinfo {author} {\bibfnamefont {D.~M.}\ \bibnamefont {Zajac}}, \bibinfo
  {author} {\bibfnamefont {J.}~\bibnamefont {Stehlik}}, \bibinfo {author}
  {\bibfnamefont {L.~F.}\ \bibnamefont {Edge}}, \ and\ \bibinfo {author}
  {\bibfnamefont {J.~R.}\ \bibnamefont {Petta}},\ }\href {\doibase
  10.1063/1.4974536} {\bibfield  {journal} {\bibinfo  {journal} {Applied
  Physics Letters}\ }\textbf {\bibinfo {volume} {110}},\ \bibinfo {pages}
  {043502} (\bibinfo {year} {2017})}\BibitemShut {NoStop}%
\bibitem [{\citenamefont {Hill}\ \emph {et~al.}(2015)\citenamefont {Hill},
  \citenamefont {Peretz}, \citenamefont {Hile}, \citenamefont {House},
  \citenamefont {Fuechsle}, \citenamefont {Rogge}, \citenamefont {Simmons},\
  and\ \citenamefont {Hollenberg}}]{hill_surface_2015}%
  \BibitemOpen
  \bibfield  {author} {\bibinfo {author} {\bibfnamefont {C.~D.}\ \bibnamefont
  {Hill}}, \bibinfo {author} {\bibfnamefont {E.}~\bibnamefont {Peretz}},
  \bibinfo {author} {\bibfnamefont {S.~J.}\ \bibnamefont {Hile}}, \bibinfo
  {author} {\bibfnamefont {M.~G.}\ \bibnamefont {House}}, \bibinfo {author}
  {\bibfnamefont {M.}~\bibnamefont {Fuechsle}}, \bibinfo {author}
  {\bibfnamefont {S.}~\bibnamefont {Rogge}}, \bibinfo {author} {\bibfnamefont
  {M.~Y.}\ \bibnamefont {Simmons}}, \ and\ \bibinfo {author} {\bibfnamefont
  {L.~C.~L.}\ \bibnamefont {Hollenberg}},\ }\href
  {http://advances.sciencemag.org/content/1/9/e1500707} {\bibfield  {journal}
  {\bibinfo  {journal} {Science Advances}\ }\textbf {\bibinfo {volume} {1}}
  (\bibinfo {year} {2015})}\BibitemShut {NoStop}%
\bibitem [{\citenamefont {Morley}(2015)}]{morley_review_2014}%
  \BibitemOpen
  \bibfield  {author} {\bibinfo {author} {\bibfnamefont {G.~W.}\ \bibnamefont
  {Morley}},\ }\bibfield  {booktitle} {\emph {\bibinfo {booktitle} {Electron
  Paramagnetic Resonance}},\ }\href {\doibase 10.1039/9781782620280-00062} {\
  \textbf {\bibinfo {volume} {24}},\ \bibinfo {pages} {62} (\bibinfo {year}
  {2015})}\BibitemShut {NoStop}%
\bibitem [{\citenamefont {DiVincenzo}\ and\ \citenamefont
  {Loss}(1998)}]{divincenzo_quantum_1998}%
  \BibitemOpen
  \bibfield  {author} {\bibinfo {author} {\bibfnamefont {D.}~\bibnamefont
  {DiVincenzo}}\ and\ \bibinfo {author} {\bibfnamefont {D.}~\bibnamefont
  {Loss}},\ }\href {\doibase 10.1006/spmi.1997.0520} {\bibfield  {journal}
  {\bibinfo  {journal} {Superlattices and Microstructures}\ }\textbf {\bibinfo
  {volume} {23}},\ \bibinfo {pages} {419 } (\bibinfo {year}
  {1998})}\BibitemShut {NoStop}%
\bibitem [{\citenamefont {Kane}(1998)}]{kane_silicon-based_1998}%
  \BibitemOpen
  \bibfield  {author} {\bibinfo {author} {\bibfnamefont {B.~E.}\ \bibnamefont
  {Kane}},\ }\href {\doibase 10.1038/30156} {\bibfield  {journal} {\bibinfo
  {journal} {Nature}\ }\textbf {\bibinfo {volume} {393}},\ \bibinfo {pages}
  {133} (\bibinfo {year} {1998})}\BibitemShut {NoStop}%
\bibitem [{\citenamefont {Morton}\ \emph {et~al.}(2008)\citenamefont {Morton},
  \citenamefont {Tyryshkin}, \citenamefont {Brown}, \citenamefont {Shankar},
  \citenamefont {Lovett}, \citenamefont {Ardavan}, \citenamefont {Schenkel},
  \citenamefont {Haller}, \citenamefont {Ager},\ and\ \citenamefont
  {Lyon}}]{morton_solid-state_2008}%
  \BibitemOpen
  \bibfield  {author} {\bibinfo {author} {\bibfnamefont {J.~J.~L.}\
  \bibnamefont {Morton}}, \bibinfo {author} {\bibfnamefont {A.~M.}\
  \bibnamefont {Tyryshkin}}, \bibinfo {author} {\bibfnamefont {R.~M.}\
  \bibnamefont {Brown}}, \bibinfo {author} {\bibfnamefont {S.}~\bibnamefont
  {Shankar}}, \bibinfo {author} {\bibfnamefont {B.~W.}\ \bibnamefont {Lovett}},
  \bibinfo {author} {\bibfnamefont {A.}~\bibnamefont {Ardavan}}, \bibinfo
  {author} {\bibfnamefont {T.}~\bibnamefont {Schenkel}}, \bibinfo {author}
  {\bibfnamefont {E.~E.}\ \bibnamefont {Haller}}, \bibinfo {author}
  {\bibfnamefont {J.~W.}\ \bibnamefont {Ager}}, \ and\ \bibinfo {author}
  {\bibfnamefont {S.~A.}\ \bibnamefont {Lyon}},\ }\href {\doibase
  10.1038/nature07295} {\bibfield  {journal} {\bibinfo  {journal} {Nature}\
  }\textbf {\bibinfo {volume} {455}},\ \bibinfo {pages} {1085} (\bibinfo {year}
  {2008})}\BibitemShut {NoStop}%
\bibitem [{\citenamefont {Pla}\ \emph {et~al.}(2013)\citenamefont {Pla},
  \citenamefont {Tan}, \citenamefont {Dehollain}, \citenamefont {Lim},
  \citenamefont {Morton}, \citenamefont {Zwanenburg}, \citenamefont {Jamieson},
  \citenamefont {Dzurak},\ and\ \citenamefont
  {Morello}}]{pla_high-fidelity_2013}%
  \BibitemOpen
  \bibfield  {author} {\bibinfo {author} {\bibfnamefont {J.~J.}\ \bibnamefont
  {Pla}}, \bibinfo {author} {\bibfnamefont {K.~Y.}\ \bibnamefont {Tan}},
  \bibinfo {author} {\bibfnamefont {J.~P.}\ \bibnamefont {Dehollain}}, \bibinfo
  {author} {\bibfnamefont {W.~H.}\ \bibnamefont {Lim}}, \bibinfo {author}
  {\bibfnamefont {J.~J.~L.}\ \bibnamefont {Morton}}, \bibinfo {author}
  {\bibfnamefont {F.~A.}\ \bibnamefont {Zwanenburg}}, \bibinfo {author}
  {\bibfnamefont {D.~N.}\ \bibnamefont {Jamieson}}, \bibinfo {author}
  {\bibfnamefont {A.~S.}\ \bibnamefont {Dzurak}}, \ and\ \bibinfo {author}
  {\bibfnamefont {A.}~\bibnamefont {Morello}},\ }\href {\doibase
  10.1038/nature12011} {\bibfield  {journal} {\bibinfo  {journal} {Nature}\
  }\textbf {\bibinfo {volume} {496}},\ \bibinfo {pages} {334} (\bibinfo {year}
  {2013})}\BibitemShut {NoStop}%
\bibitem [{\citenamefont {Saeedi}\ \emph {et~al.}(2013)\citenamefont {Saeedi},
  \citenamefont {Simmons}, \citenamefont {Salvail}, \citenamefont {Dluhy},
  \citenamefont {Riemann}, \citenamefont {Abrosimov}, \citenamefont {Becker},
  \citenamefont {Pohl}, \citenamefont {Morton},\ and\ \citenamefont
  {Thewalt}}]{saeedi_room-temperature_2013}%
  \BibitemOpen
  \bibfield  {author} {\bibinfo {author} {\bibfnamefont {K.}~\bibnamefont
  {Saeedi}}, \bibinfo {author} {\bibfnamefont {S.}~\bibnamefont {Simmons}},
  \bibinfo {author} {\bibfnamefont {J.~Z.}\ \bibnamefont {Salvail}}, \bibinfo
  {author} {\bibfnamefont {P.}~\bibnamefont {Dluhy}}, \bibinfo {author}
  {\bibfnamefont {H.}~\bibnamefont {Riemann}}, \bibinfo {author} {\bibfnamefont
  {N.~V.}\ \bibnamefont {Abrosimov}}, \bibinfo {author} {\bibfnamefont
  {P.}~\bibnamefont {Becker}}, \bibinfo {author} {\bibfnamefont {H.-J.}\
  \bibnamefont {Pohl}}, \bibinfo {author} {\bibfnamefont {J.~J.~L.}\
  \bibnamefont {Morton}}, \ and\ \bibinfo {author} {\bibfnamefont {M.~L.~W.}\
  \bibnamefont {Thewalt}},\ }\href {\doibase 10.1126/science.1239584}
  {\bibfield  {journal} {\bibinfo  {journal} {Science}\ }\textbf {\bibinfo
  {volume} {342}},\ \bibinfo {pages} {830} (\bibinfo {year}
  {2013})}\BibitemShut {NoStop}%
\bibitem [{\citenamefont {Saeedi}(2014)}]{saeedi_optical_2014}%
  \BibitemOpen
  \bibfield  {author} {\bibinfo {author} {\bibfnamefont {I.~K.}\ \bibnamefont
  {Saeedi}},\ }\emph {\bibinfo {title} {Optical NMR Study of 31P Donor Spins in
  Isotopically Enriched 28Si}},\ \href@noop {} {Ph.D. thesis},\ \bibinfo
  {school} {Simon Fraser University} (\bibinfo {year} {2014})\BibitemShut
  {NoStop}%
\bibitem [{\citenamefont {Morello}\ \emph {et~al.}(2010)\citenamefont
  {Morello}, \citenamefont {Pla}, \citenamefont {Zwanenburg}, \citenamefont
  {Chan}, \citenamefont {Tan}, \citenamefont {Huebl}, \citenamefont
  {M\"ott\"nen}, \citenamefont {Nugroho}, \citenamefont {Yang}, \citenamefont
  {van Donkelaar}, \citenamefont {Alves}, \citenamefont {Jamieson},
  \citenamefont {Escott}, \citenamefont {Hollenberg}, \citenamefont {Clark},\
  and\ \citenamefont {Dzurak}}]{morello_single-shot_2010}%
  \BibitemOpen
  \bibfield  {author} {\bibinfo {author} {\bibfnamefont {A.}~\bibnamefont
  {Morello}}, \bibinfo {author} {\bibfnamefont {J.~J.}\ \bibnamefont {Pla}},
  \bibinfo {author} {\bibfnamefont {F.~A.}\ \bibnamefont {Zwanenburg}},
  \bibinfo {author} {\bibfnamefont {K.~W.}\ \bibnamefont {Chan}}, \bibinfo
  {author} {\bibfnamefont {K.~Y.}\ \bibnamefont {Tan}}, \bibinfo {author}
  {\bibfnamefont {H.}~\bibnamefont {Huebl}}, \bibinfo {author} {\bibfnamefont
  {M.}~\bibnamefont {M\"ott\"nen}}, \bibinfo {author} {\bibfnamefont {C.~D.}\
  \bibnamefont {Nugroho}}, \bibinfo {author} {\bibfnamefont {C.}~\bibnamefont
  {Yang}}, \bibinfo {author} {\bibfnamefont {J.~A.}\ \bibnamefont {van
  Donkelaar}}, \bibinfo {author} {\bibfnamefont {A.~D.~C.}\ \bibnamefont
  {Alves}}, \bibinfo {author} {\bibfnamefont {D.~N.}\ \bibnamefont {Jamieson}},
  \bibinfo {author} {\bibfnamefont {C.~C.}\ \bibnamefont {Escott}}, \bibinfo
  {author} {\bibfnamefont {L.~C.~L.}\ \bibnamefont {Hollenberg}}, \bibinfo
  {author} {\bibfnamefont {R.~G.}\ \bibnamefont {Clark}}, \ and\ \bibinfo
  {author} {\bibfnamefont {A.~S.}\ \bibnamefont {Dzurak}},\ }\href {\doibase
  10.1038/nature09392} {\bibfield  {journal} {\bibinfo  {journal} {Nature}\
  }\textbf {\bibinfo {volume} {467}},\ \bibinfo {pages} {687} (\bibinfo {year}
  {2010})}\BibitemShut {NoStop}%
\bibitem [{\citenamefont {Fuechsle}\ \emph {et~al.}(2012)\citenamefont
  {Fuechsle}, \citenamefont {Miwa}, \citenamefont {Mahapatra}, \citenamefont
  {Ryu}, \citenamefont {Lee}, \citenamefont {Warschkow}, \citenamefont
  {Hollenberg}, \citenamefont {Klimeck},\ and\ \citenamefont
  {Simmons}}]{fuechsle_single-atom_2012}%
  \BibitemOpen
  \bibfield  {author} {\bibinfo {author} {\bibfnamefont {M.}~\bibnamefont
  {Fuechsle}}, \bibinfo {author} {\bibfnamefont {J.~A.}\ \bibnamefont {Miwa}},
  \bibinfo {author} {\bibfnamefont {S.}~\bibnamefont {Mahapatra}}, \bibinfo
  {author} {\bibfnamefont {H.}~\bibnamefont {Ryu}}, \bibinfo {author}
  {\bibfnamefont {S.}~\bibnamefont {Lee}}, \bibinfo {author} {\bibfnamefont
  {O.}~\bibnamefont {Warschkow}}, \bibinfo {author} {\bibfnamefont {L.~C.~L.}\
  \bibnamefont {Hollenberg}}, \bibinfo {author} {\bibfnamefont
  {G.}~\bibnamefont {Klimeck}}, \ and\ \bibinfo {author} {\bibfnamefont
  {M.~Y.}\ \bibnamefont {Simmons}},\ }\href {\doibase 10.1038/nnano.2012.21}
  {\bibfield  {journal} {\bibinfo  {journal} {Nature Nanotechnology}\ }\textbf
  {\bibinfo {volume} {7}},\ \bibinfo {pages} {242} (\bibinfo {year}
  {2012})}\BibitemShut {NoStop}%
\bibitem [{\citenamefont {Lo}\ \emph {et~al.}(2015)\citenamefont {Lo},
  \citenamefont {Urdampilleta}, \citenamefont {Ross}, \citenamefont
  {Gonzalez-Zalba}, \citenamefont {Mansir}, \citenamefont {Lyon}, \citenamefont
  {Thewalt},\ and\ \citenamefont {Morton}}]{lo_hybrid_2015}%
  \BibitemOpen
  \bibfield  {author} {\bibinfo {author} {\bibfnamefont {C.~C.}\ \bibnamefont
  {Lo}}, \bibinfo {author} {\bibfnamefont {M.}~\bibnamefont {Urdampilleta}},
  \bibinfo {author} {\bibfnamefont {P.}~\bibnamefont {Ross}}, \bibinfo {author}
  {\bibfnamefont {M.~F.}\ \bibnamefont {Gonzalez-Zalba}}, \bibinfo {author}
  {\bibfnamefont {J.}~\bibnamefont {Mansir}}, \bibinfo {author} {\bibfnamefont
  {S.~A.}\ \bibnamefont {Lyon}}, \bibinfo {author} {\bibfnamefont {M.~L.~W.}\
  \bibnamefont {Thewalt}}, \ and\ \bibinfo {author} {\bibfnamefont {J.~J.~L.}\
  \bibnamefont {Morton}},\ }\href {\doibase 10.1038/nmat4250} {\bibfield
  {journal} {\bibinfo  {journal} {Nature Materials}\ }\textbf {\bibinfo
  {volume} {14}},\ \bibinfo {pages} {490} (\bibinfo {year} {2015})}\BibitemShut
  {NoStop}%
\bibitem [{\citenamefont {Saeedi}\ \emph {et~al.}(2015)\citenamefont {Saeedi},
  \citenamefont {Szech}, \citenamefont {Dluhy}, \citenamefont {Salvail},
  \citenamefont {Morse}, \citenamefont {Riemann}, \citenamefont {Abrosimov},
  \citenamefont {Notzel}, \citenamefont {Litvinenko}, \citenamefont {Murdin},\
  and\ \citenamefont {Thewalt}}]{saeedi_donorbound_2015}%
  \BibitemOpen
  \bibfield  {author} {\bibinfo {author} {\bibfnamefont {K.}~\bibnamefont
  {Saeedi}}, \bibinfo {author} {\bibfnamefont {M.}~\bibnamefont {Szech}},
  \bibinfo {author} {\bibfnamefont {P.}~\bibnamefont {Dluhy}}, \bibinfo
  {author} {\bibfnamefont {J.~Z.}\ \bibnamefont {Salvail}}, \bibinfo {author}
  {\bibfnamefont {K.~J.}\ \bibnamefont {Morse}}, \bibinfo {author}
  {\bibfnamefont {H.}~\bibnamefont {Riemann}}, \bibinfo {author} {\bibfnamefont
  {N.~V.}\ \bibnamefont {Abrosimov}}, \bibinfo {author} {\bibfnamefont
  {N.}~\bibnamefont {Notzel}}, \bibinfo {author} {\bibfnamefont {K.~L.}\
  \bibnamefont {Litvinenko}}, \bibinfo {author} {\bibfnamefont {B.~N.}\
  \bibnamefont {Murdin}}, \ and\ \bibinfo {author} {\bibfnamefont {M.~L.~W.}\
  \bibnamefont {Thewalt}},\ }\href {\doibase 10.1038/srep10493} {\bibfield
  {journal} {\bibinfo  {journal} {Scientific Reports}\ }\textbf {\bibinfo
  {volume} {5}},\ \bibinfo {pages} {10493} (\bibinfo {year}
  {2015})}\BibitemShut {NoStop}%
\bibitem [{\citenamefont {Lloyd}(1996)}]{lloyd_universal_1996}%
  \BibitemOpen
  \bibfield  {author} {\bibinfo {author} {\bibfnamefont {S.}~\bibnamefont
  {Lloyd}},\ }\href {\doibase 10.1126/science.273.5278.1073} {\bibfield
  {journal} {\bibinfo  {journal} {Science}\ }\textbf {\bibinfo {volume}
  {273}},\ \bibinfo {pages} {1073} (\bibinfo {year} {1996})}\BibitemShut
  {NoStop}%
\bibitem [{\citenamefont {Koiller}\ \emph {et~al.}(2001)\citenamefont
  {Koiller}, \citenamefont {Hu},\ and\ \citenamefont
  {Das~Sarma}}]{koiller_exchange_2001}%
  \BibitemOpen
  \bibfield  {author} {\bibinfo {author} {\bibfnamefont {B.}~\bibnamefont
  {Koiller}}, \bibinfo {author} {\bibfnamefont {X.}~\bibnamefont {Hu}}, \ and\
  \bibinfo {author} {\bibfnamefont {S.}~\bibnamefont {Das~Sarma}},\ }\href
  {\doibase 10.1103/PhysRevLett.88.027903} {\bibfield  {journal} {\bibinfo
  {journal} {Physical Review Letters}\ }\textbf {\bibinfo {volume} {88}},\
  \bibinfo {pages} {027903} (\bibinfo {year} {2001})}\BibitemShut {NoStop}%
\bibitem [{\citenamefont {Hersam}\ \emph {et~al.}(2000)\citenamefont {Hersam},
  \citenamefont {Guisinger},\ and\ \citenamefont
  {Lyding}}]{hersam_silicon-based_2000}%
  \BibitemOpen
  \bibfield  {author} {\bibinfo {author} {\bibfnamefont {M.~C.}\ \bibnamefont
  {Hersam}}, \bibinfo {author} {\bibfnamefont {N.~P.}\ \bibnamefont
  {Guisinger}}, \ and\ \bibinfo {author} {\bibfnamefont {J.~W.}\ \bibnamefont
  {Lyding}},\ }\href {\doibase 10.1088/0957-4484/11/2/306} {\bibfield
  {journal} {\bibinfo  {journal} {Nanotechnology}\ }\textbf {\bibinfo {volume}
  {11}},\ \bibinfo {pages} {70} (\bibinfo {year} {2000})}\BibitemShut {NoStop}%
\bibitem [{\citenamefont {O’Brien}\ \emph {et~al.}(2002)\citenamefont
  {O’Brien}, \citenamefont {Schofield}, \citenamefont {Simmons},
  \citenamefont {Clark}, \citenamefont {Dzurak}, \citenamefont {Curson},
  \citenamefont {Kane}, \citenamefont {McAlpine}, \citenamefont {Hawley},\ and\
  \citenamefont {Brown}}]{obrien_scanning_2002}%
  \BibitemOpen
  \bibfield  {author} {\bibinfo {author} {\bibfnamefont {J.~L.}\ \bibnamefont
  {O’Brien}}, \bibinfo {author} {\bibfnamefont {S.~R.}\ \bibnamefont
  {Schofield}}, \bibinfo {author} {\bibfnamefont {M.~Y.}\ \bibnamefont
  {Simmons}}, \bibinfo {author} {\bibfnamefont {R.~G.}\ \bibnamefont {Clark}},
  \bibinfo {author} {\bibfnamefont {A.~S.}\ \bibnamefont {Dzurak}}, \bibinfo
  {author} {\bibfnamefont {N.~J.}\ \bibnamefont {Curson}}, \bibinfo {author}
  {\bibfnamefont {B.~E.}\ \bibnamefont {Kane}}, \bibinfo {author}
  {\bibfnamefont {N.~S.}\ \bibnamefont {McAlpine}}, \bibinfo {author}
  {\bibfnamefont {M.~E.}\ \bibnamefont {Hawley}}, \ and\ \bibinfo {author}
  {\bibfnamefont {G.~W.}\ \bibnamefont {Brown}},\ }\href {\doibase
  10.1088/0964-1726/11/5/318} {\bibfield  {journal} {\bibinfo  {journal} {Smart
  Materials and Structures}\ }\textbf {\bibinfo {volume} {11}},\ \bibinfo
  {pages} {741} (\bibinfo {year} {2002})}\BibitemShut {NoStop}%
\bibitem [{\citenamefont {Broome}\ \emph {et~al.}(2018)\citenamefont {Broome},
  \citenamefont {Gorman}, \citenamefont {House}, \citenamefont {Hile},
  \citenamefont {Keizer}, \citenamefont {Keith}, \citenamefont {Hill},
  \citenamefont {Watson}, \citenamefont {Baker}, \citenamefont {Hollenberg},\
  and\ \citenamefont {Simmons}}]{broome_two-electron_2018}%
  \BibitemOpen
  \bibfield  {author} {\bibinfo {author} {\bibfnamefont {M.~A.}\ \bibnamefont
  {Broome}}, \bibinfo {author} {\bibfnamefont {S.~K.}\ \bibnamefont {Gorman}},
  \bibinfo {author} {\bibfnamefont {M.~G.}\ \bibnamefont {House}}, \bibinfo
  {author} {\bibfnamefont {S.~J.}\ \bibnamefont {Hile}}, \bibinfo {author}
  {\bibfnamefont {J.~G.}\ \bibnamefont {Keizer}}, \bibinfo {author}
  {\bibfnamefont {D.}~\bibnamefont {Keith}}, \bibinfo {author} {\bibfnamefont
  {C.~D.}\ \bibnamefont {Hill}}, \bibinfo {author} {\bibfnamefont {T.~F.}\
  \bibnamefont {Watson}}, \bibinfo {author} {\bibfnamefont {W.~J.}\
  \bibnamefont {Baker}}, \bibinfo {author} {\bibfnamefont {L.~C.~L.}\
  \bibnamefont {Hollenberg}}, \ and\ \bibinfo {author} {\bibfnamefont {M.~Y.}\
  \bibnamefont {Simmons}},\ }\href {\doibase 10.1038/s41467-018-02982-x}
  {\bibfield  {journal} {\bibinfo  {journal} {Nature Communications}\ }\textbf
  {\bibinfo {volume} {9}},\ \bibinfo {pages} {980} (\bibinfo {year}
  {2018})}\BibitemShut {NoStop}%
\bibitem [{\citenamefont {Tosi}\ \emph {et~al.}(2017)\citenamefont {Tosi},
  \citenamefont {Mohiyaddin}, \citenamefont {Schmitt}, \citenamefont {Tenberg},
  \citenamefont {Rahman}, \citenamefont {Klimeck},\ and\ \citenamefont
  {Morello}}]{tosi_silicon_2017}%
  \BibitemOpen
  \bibfield  {author} {\bibinfo {author} {\bibfnamefont {G.}~\bibnamefont
  {Tosi}}, \bibinfo {author} {\bibfnamefont {F.~A.}\ \bibnamefont
  {Mohiyaddin}}, \bibinfo {author} {\bibfnamefont {V.}~\bibnamefont {Schmitt}},
  \bibinfo {author} {\bibfnamefont {S.}~\bibnamefont {Tenberg}}, \bibinfo
  {author} {\bibfnamefont {R.}~\bibnamefont {Rahman}}, \bibinfo {author}
  {\bibfnamefont {G.}~\bibnamefont {Klimeck}}, \ and\ \bibinfo {author}
  {\bibfnamefont {A.}~\bibnamefont {Morello}},\ }\href {\doibase
  10.1038/s41467-017-00378-x} {\bibfield  {journal} {\bibinfo  {journal}
  {Nature Communications}\ }\textbf {\bibinfo {volume} {8}},\ \bibinfo {pages}
  {450} (\bibinfo {year} {2017})}\BibitemShut {NoStop}%
\bibitem [{\citenamefont {Stoneham}\ \emph {et~al.}(2003)\citenamefont
  {Stoneham}, \citenamefont {Fisher},\ and\ \citenamefont
  {Greenland}}]{stoneham_optically_2003}%
  \BibitemOpen
  \bibfield  {author} {\bibinfo {author} {\bibfnamefont {A.~M.}\ \bibnamefont
  {Stoneham}}, \bibinfo {author} {\bibfnamefont {A.~J.}\ \bibnamefont
  {Fisher}}, \ and\ \bibinfo {author} {\bibfnamefont {P.~T.}\ \bibnamefont
  {Greenland}},\ }\href {\doibase 10.1088/0953-8984/15/27/102} {\bibfield
  {journal} {\bibinfo  {journal} {Journal of Physics: Condensed Matter}\
  }\textbf {\bibinfo {volume} {15}},\ \bibinfo {pages} {L447} (\bibinfo {year}
  {2003})}\BibitemShut {NoStop}%
\bibitem [{\citenamefont {Greenland}\ \emph {et~al.}(2010)\citenamefont
  {Greenland}, \citenamefont {Lynch}, \citenamefont {van~der Meer},
  \citenamefont {Murdin}, \citenamefont {Pidgeon}, \citenamefont {Redlich},
  \citenamefont {Vinh},\ and\ \citenamefont
  {Aeppli}}]{greenland_coherent_2010}%
  \BibitemOpen
  \bibfield  {author} {\bibinfo {author} {\bibfnamefont {P.~T.}\ \bibnamefont
  {Greenland}}, \bibinfo {author} {\bibfnamefont {S.~a.}\ \bibnamefont
  {Lynch}}, \bibinfo {author} {\bibfnamefont {a.~F.~G.}\ \bibnamefont {van~der
  Meer}}, \bibinfo {author} {\bibfnamefont {B.~N.}\ \bibnamefont {Murdin}},
  \bibinfo {author} {\bibfnamefont {C.~R.}\ \bibnamefont {Pidgeon}}, \bibinfo
  {author} {\bibfnamefont {B.}~\bibnamefont {Redlich}}, \bibinfo {author}
  {\bibfnamefont {N.~Q.}\ \bibnamefont {Vinh}}, \ and\ \bibinfo {author}
  {\bibfnamefont {G.}~\bibnamefont {Aeppli}},\ }\href
  {https://www.nature.com/articles/nature09112} {\bibfield  {journal} {\bibinfo
   {journal} {Nature}\ }\textbf {\bibinfo {volume} {465}},\ \bibinfo {pages}
  {1057} (\bibinfo {year} {2010})}\BibitemShut {NoStop}%
\bibitem [{\citenamefont {Stoneham}(2009)}]{stoneham_room-temperature_2009}%
  \BibitemOpen
  \bibfield  {author} {\bibinfo {author} {\bibfnamefont {M.}~\bibnamefont
  {Stoneham}},\ }\href {https://physics.aps.org/articles/v2/34} {\bibfield
  {journal} {\bibinfo  {journal} {Physics}\ }\textbf {\bibinfo {volume} {2}},\
  \bibinfo {pages} {34} (\bibinfo {year} {2009})}\BibitemShut {NoStop}%
\bibitem [{\citenamefont {Rodriquez}\ \emph {et~al.}(2004)\citenamefont
  {Rodriquez}, \citenamefont {Fisher}, \citenamefont {Greenland},\ and\
  \citenamefont {Stoneham}}]{rodriquez_avoiding_2004}%
  \BibitemOpen
  \bibfield  {author} {\bibinfo {author} {\bibfnamefont {R.}~\bibnamefont
  {Rodriquez}}, \bibinfo {author} {\bibfnamefont {A.~J.}\ \bibnamefont
  {Fisher}}, \bibinfo {author} {\bibfnamefont {P.~T.}\ \bibnamefont
  {Greenland}}, \ and\ \bibinfo {author} {\bibfnamefont {A.~M.}\ \bibnamefont
  {Stoneham}},\ }\href {\doibase 10.1088/0953-8984/16/16/001} {\bibfield
  {journal} {\bibinfo  {journal} {Journal of Physics: Condensed Matter}\
  }\textbf {\bibinfo {volume} {16}},\ \bibinfo {pages} {2757} (\bibinfo {year}
  {2004})}\BibitemShut {NoStop}%
\bibitem [{\citenamefont {Morse}\ \emph {et~al.}(2018)\citenamefont {Morse},
  \citenamefont {Dluhy}, \citenamefont {Huber}, \citenamefont {Salvail},
  \citenamefont {Saeedi}, \citenamefont {Riemann}, \citenamefont {Abrosimov},
  \citenamefont {Becker}, \citenamefont {Pohl}, \citenamefont {Simmons},\ and\
  \citenamefont {Thewalt}}]{morse_zero-field_2018}%
  \BibitemOpen
  \bibfield  {author} {\bibinfo {author} {\bibfnamefont {K.~J.}\ \bibnamefont
  {Morse}}, \bibinfo {author} {\bibfnamefont {P.}~\bibnamefont {Dluhy}},
  \bibinfo {author} {\bibfnamefont {J.}~\bibnamefont {Huber}}, \bibinfo
  {author} {\bibfnamefont {J.~Z.}\ \bibnamefont {Salvail}}, \bibinfo {author}
  {\bibfnamefont {K.}~\bibnamefont {Saeedi}}, \bibinfo {author} {\bibfnamefont
  {H.}~\bibnamefont {Riemann}}, \bibinfo {author} {\bibfnamefont {N.~V.}\
  \bibnamefont {Abrosimov}}, \bibinfo {author} {\bibfnamefont {P.}~\bibnamefont
  {Becker}}, \bibinfo {author} {\bibfnamefont {H.-J.}\ \bibnamefont {Pohl}},
  \bibinfo {author} {\bibfnamefont {S.}~\bibnamefont {Simmons}}, \ and\
  \bibinfo {author} {\bibfnamefont {M.~L.~W.}\ \bibnamefont {Thewalt}},\ }\href
  {https://journals.aps.org/prb/abstract/10.1103/PhysRevB.97.115205} {\bibfield
   {journal} {\bibinfo  {journal} {Physical Review B}\ }\textbf {\bibinfo
  {volume} {97}} (\bibinfo {year} {2018})}\BibitemShut {NoStop}%
\bibitem [{\citenamefont {Stockbridge}\ \emph {et~al.}()\citenamefont
  {Stockbridge}, \citenamefont {Chick}, \citenamefont {Crane}, \citenamefont
  {Fisher},\ and\ \citenamefont {Murdin}}]{paper2}%
  \BibitemOpen
  \bibfield  {author} {\bibinfo {author} {\bibfnamefont {K.}~\bibnamefont
  {Stockbridge}}, \bibinfo {author} {\bibfnamefont {S.}~\bibnamefont {Chick}},
  \bibinfo {author} {\bibfnamefont {E.}~\bibnamefont {Crane}}, \bibinfo
  {author} {\bibfnamefont {A.}~\bibnamefont {Fisher}}, \ and\ \bibinfo {author}
  {\bibfnamefont {B.}~\bibnamefont {Murdin}},\ }\href@noop {} {\bibinfo
  {journal} {in preparation}\ }\BibitemShut {NoStop}%
\bibitem [{\citenamefont {Hazzard}\ \emph {et~al.}(2014)\citenamefont
  {Hazzard}, \citenamefont {Gadway}, \citenamefont {Foss-Feig}, \citenamefont
  {Yan}, \citenamefont {Moses}, \citenamefont {Covey}, \citenamefont {Yao},
  \citenamefont {Lukin}, \citenamefont {Ye}, \citenamefont {Jin},\ and\
  \citenamefont {Rey}}]{hazzard_many-body_2014}%
  \BibitemOpen
\bibfield  {journal} {  }\bibfield  {author} {\bibinfo {author} {\bibfnamefont
  {K.~R.~A.}\ \bibnamefont {Hazzard}}, \bibinfo {author} {\bibfnamefont
  {B.}~\bibnamefont {Gadway}}, \bibinfo {author} {\bibfnamefont
  {M.}~\bibnamefont {Foss-Feig}}, \bibinfo {author} {\bibfnamefont
  {B.}~\bibnamefont {Yan}}, \bibinfo {author} {\bibfnamefont {S.~A.}\
  \bibnamefont {Moses}}, \bibinfo {author} {\bibfnamefont {J.~P.}\ \bibnamefont
  {Covey}}, \bibinfo {author} {\bibfnamefont {N.~Y.}\ \bibnamefont {Yao}},
  \bibinfo {author} {\bibfnamefont {M.~D.}\ \bibnamefont {Lukin}}, \bibinfo
  {author} {\bibfnamefont {J.}~\bibnamefont {Ye}}, \bibinfo {author}
  {\bibfnamefont {D.~S.}\ \bibnamefont {Jin}}, \ and\ \bibinfo {author}
  {\bibfnamefont {A.~M.}\ \bibnamefont {Rey}},\ }\href {\doibase
  10.1103/PhysRevLett.113.195302} {\bibfield  {journal} {\bibinfo  {journal}
  {Phys. Rev. Lett.}\ }\textbf {\bibinfo {volume} {113}},\ \bibinfo {pages}
  {195302} (\bibinfo {year} {2014})}\BibitemShut {NoStop}%
\bibitem [{\citenamefont {Vinh}\ \emph {et~al.}(2008)\citenamefont {Vinh},
  \citenamefont {Greenland}, \citenamefont {Litvinenko}, \citenamefont
  {Redlich}, \citenamefont {van~der Meer}, \citenamefont {Lynch}, \citenamefont
  {Warner}, \citenamefont {Stoneham}, \citenamefont {Aeppli}, \citenamefont
  {Paul}, \citenamefont {Pidgeon},\ and\ \citenamefont
  {Murdin}}]{vinh_silicon_2008}%
  \BibitemOpen
  \bibfield  {author} {\bibinfo {author} {\bibfnamefont {N.~Q.}\ \bibnamefont
  {Vinh}}, \bibinfo {author} {\bibfnamefont {P.~T.}\ \bibnamefont {Greenland}},
  \bibinfo {author} {\bibfnamefont {K.}~\bibnamefont {Litvinenko}}, \bibinfo
  {author} {\bibfnamefont {B.}~\bibnamefont {Redlich}}, \bibinfo {author}
  {\bibfnamefont {A.~F.~G.}\ \bibnamefont {van~der Meer}}, \bibinfo {author}
  {\bibfnamefont {S.~A.}\ \bibnamefont {Lynch}}, \bibinfo {author}
  {\bibfnamefont {M.}~\bibnamefont {Warner}}, \bibinfo {author} {\bibfnamefont
  {A.~M.}\ \bibnamefont {Stoneham}}, \bibinfo {author} {\bibfnamefont
  {G.}~\bibnamefont {Aeppli}}, \bibinfo {author} {\bibfnamefont {D.~J.}\
  \bibnamefont {Paul}}, \bibinfo {author} {\bibfnamefont {C.~R.}\ \bibnamefont
  {Pidgeon}}, \ and\ \bibinfo {author} {\bibfnamefont {B.~N.}\ \bibnamefont
  {Murdin}},\ }\href {\doibase 10.1073/pnas.0802721105} {\bibfield  {journal}
  {\bibinfo  {journal} {Proceedings of the National Academy of Sciences}\
  }\textbf {\bibinfo {volume} {105}},\ \bibinfo {pages} {10649} (\bibinfo
  {year} {2008})}\BibitemShut {NoStop}%
\bibitem [{\citenamefont {Litvinenko}\ \emph {et~al.}(2015)\citenamefont
  {Litvinenko}, \citenamefont {Bowyer}, \citenamefont {Greenland},
  \citenamefont {Stavrias}, \citenamefont {Li}, \citenamefont {Gwilliam},
  \citenamefont {Villis}, \citenamefont {Matmon}, \citenamefont {Pang},
  \citenamefont {Redlich}, \citenamefont {a.F.G. van~der Meer}, \citenamefont
  {Pidgeon}, \citenamefont {Aeppli},\ and\ \citenamefont
  {Murdin}}]{litvinenko_coherent_2015}%
  \BibitemOpen
  \bibfield  {author} {\bibinfo {author} {\bibfnamefont {K.}~\bibnamefont
  {Litvinenko}}, \bibinfo {author} {\bibfnamefont {E.}~\bibnamefont {Bowyer}},
  \bibinfo {author} {\bibfnamefont {P.}~\bibnamefont {Greenland}}, \bibinfo
  {author} {\bibfnamefont {N.}~\bibnamefont {Stavrias}}, \bibinfo {author}
  {\bibfnamefont {J.}~\bibnamefont {Li}}, \bibinfo {author} {\bibfnamefont
  {R.}~\bibnamefont {Gwilliam}}, \bibinfo {author} {\bibfnamefont
  {B.}~\bibnamefont {Villis}}, \bibinfo {author} {\bibfnamefont
  {G.}~\bibnamefont {Matmon}}, \bibinfo {author} {\bibfnamefont
  {M.}~\bibnamefont {Pang}}, \bibinfo {author} {\bibfnamefont {B.}~\bibnamefont
  {Redlich}}, \bibinfo {author} {\bibnamefont {a.F.G. van~der Meer}}, \bibinfo
  {author} {\bibfnamefont {C.}~\bibnamefont {Pidgeon}}, \bibinfo {author}
  {\bibfnamefont {G.}~\bibnamefont {Aeppli}}, \ and\ \bibinfo {author}
  {\bibfnamefont {B.}~\bibnamefont {Murdin}},\ }\href {\doibase
  10.1038/ncomms7549} {\bibfield  {journal} {\bibinfo  {journal} {Nature
  Communications}\ }\textbf {\bibinfo {volume} {6}},\ \bibinfo {pages} {6549}
  (\bibinfo {year} {2015})}\BibitemShut {NoStop}%
\bibitem [{\citenamefont {Wu}\ \emph {et~al.}(2018)\citenamefont {Wu},
  \citenamefont {Greenland}, \citenamefont {Fisher}, \citenamefont {Le},
  \citenamefont {Chick},\ and\ \citenamefont {Murdin}}]{wu_excited_2018}%
  \BibitemOpen
  \bibfield  {author} {\bibinfo {author} {\bibfnamefont {W.}~\bibnamefont
  {Wu}}, \bibinfo {author} {\bibfnamefont {P.~T.}\ \bibnamefont {Greenland}},
  \bibinfo {author} {\bibfnamefont {A.~J.}\ \bibnamefont {Fisher}}, \bibinfo
  {author} {\bibfnamefont {N.~H.}\ \bibnamefont {Le}}, \bibinfo {author}
  {\bibfnamefont {S.}~\bibnamefont {Chick}}, \ and\ \bibinfo {author}
  {\bibfnamefont {B.~N.}\ \bibnamefont {Murdin}},\ }\href
  {https://journals.aps.org/prb/abstract/10.1103/PhysRevB.97.035205} {\bibfield
   {journal} {\bibinfo  {journal} {Physical Review B}\ }\textbf {\bibinfo
  {volume} {97}},\ \bibinfo {pages} {035205} (\bibinfo {year}
  {2018})}\BibitemShut {NoStop}%
\bibitem [{\citenamefont {Levy}(2002)}]{levy_universal_2002}%
  \BibitemOpen
  \bibfield  {author} {\bibinfo {author} {\bibfnamefont {J.}~\bibnamefont
  {Levy}},\ }\href {https://doi.org/10.1103/PhysRevLett.89.147902} {\bibfield
  {journal} {\bibinfo  {journal} {Physical Review Letters}\ }\textbf {\bibinfo
  {volume} {89}},\ \bibinfo {pages} {147902} (\bibinfo {year}
  {2002})}\BibitemShut {NoStop}%
\bibitem [{\citenamefont {Keyes}(2005)}]{keyes_challenges_2005}%
  \BibitemOpen
  \bibfield  {author} {\bibinfo {author} {\bibfnamefont {R.~W.}\ \bibnamefont
  {Keyes}},\ }\href {\doibase 10.1109/MC.2005.13} {\bibfield  {journal}
  {\bibinfo  {journal} {Computer}\ }\textbf {\bibinfo {volume} {38}},\ \bibinfo
  {pages} {65} (\bibinfo {year} {2005})}\BibitemShut {NoStop}%
\bibitem [{\citenamefont {Wu}\ and\ \citenamefont
  {Fisher}(2008)}]{wu_exchange_2008}%
  \BibitemOpen
  \bibfield  {author} {\bibinfo {author} {\bibfnamefont {W.}~\bibnamefont
  {Wu}}\ and\ \bibinfo {author} {\bibfnamefont {A.~J.}\ \bibnamefont
  {Fisher}},\ }\href {\doibase 10.1103/PhysRevB.77.045201} {\bibfield
  {journal} {\bibinfo  {journal} {Phys. Rev. B}\ }\textbf {\bibinfo {volume}
  {77}},\ \bibinfo {pages} {045201} (\bibinfo {year} {2008})}\BibitemShut
  {NoStop}%
\bibitem [{\citenamefont {Trappmann}\ \emph {et~al.}(1997)\citenamefont
  {Trappmann}, \citenamefont {Surgers},\ and\ \citenamefont
  {Lohneysen}}]{trappmann_observation_1997}%
  \BibitemOpen
  \bibfield  {author} {\bibinfo {author} {\bibfnamefont {T.}~\bibnamefont
  {Trappmann}}, \bibinfo {author} {\bibfnamefont {C.}~\bibnamefont {Surgers}},
  \ and\ \bibinfo {author} {\bibfnamefont {H.}~\bibnamefont {Lohneysen}},\
  }\href {\doibase 10.1209/epl/i1997-00222-0} {\bibfield  {journal} {\bibinfo
  {journal} {Europhysics Letters (EPL)}\ }\textbf {\bibinfo {volume} {38}},\
  \bibinfo {pages} {177} (\bibinfo {year} {1997})}\BibitemShut {NoStop}%
\bibitem [{\citenamefont {Torquato}\ \emph {et~al.}(1990)\citenamefont
  {Torquato}, \citenamefont {Lu},\ and\ \citenamefont
  {Rubinstein}}]{torquato_nearest-neighbour_1989}%
  \BibitemOpen
  \bibfield  {author} {\bibinfo {author} {\bibfnamefont {S.}~\bibnamefont
  {Torquato}}, \bibinfo {author} {\bibfnamefont {B.}~\bibnamefont {Lu}}, \ and\
  \bibinfo {author} {\bibfnamefont {J.}~\bibnamefont {Rubinstein}},\ }\href
  {\doibase 10.1088/0305-4470/23/3/005} {\bibfield  {journal} {\bibinfo
  {journal} {Journal of Physics A: Mathematical and General}\ }\textbf
  {\bibinfo {volume} {23}},\ \bibinfo {pages} {L103} (\bibinfo {year}
  {1990})}\BibitemShut {NoStop}%
\bibitem [{\citenamefont {Pickard}(1982)}]{pickard_isolated_1982}%
  \BibitemOpen
  \bibfield  {author} {\bibinfo {author} {\bibfnamefont {D.~K.}\ \bibnamefont
  {Pickard}},\ }\href {\doibase 10.2307/3213499} {\bibfield  {journal}
  {\bibinfo  {journal} {Journal of Applied Probability}\ }\textbf {\bibinfo
  {volume} {19}},\ \bibinfo {pages} {444} (\bibinfo {year} {1982})}\BibitemShut
  {NoStop}%
\bibitem [{\citenamefont {Cox}(1981)}]{cox_reflexive_1981}%
  \BibitemOpen
  \bibfield  {author} {\bibinfo {author} {\bibfnamefont {T.~F.}\ \bibnamefont
  {Cox}},\ }\href {\doibase 10.2307/2530424} {\bibfield  {journal} {\bibinfo
  {journal} {Biometrics}\ }\textbf {\bibinfo {volume} {37}},\ \bibinfo {pages}
  {367} (\bibinfo {year} {1981})}\BibitemShut {NoStop}%
\bibitem [{\citenamefont {Dacey}(2010)}]{dacey_proportion_2010}%
  \BibitemOpen
  \bibfield  {author} {\bibinfo {author} {\bibfnamefont {M.~F.}\ \bibnamefont
  {Dacey}},\ }\href {\doibase 10.1111/j.1538-4632.1969.tb00632.x} {\bibfield
  {journal} {\bibinfo  {journal} {Geographical Analysis}\ }\textbf {\bibinfo
  {volume} {1}},\ \bibinfo {pages} {385} (\bibinfo {year} {2010})}\BibitemShut
  {NoStop}%
\bibitem [{\citenamefont {Bahcall J.~N.}(1981)}]{bahcall_distribution_1981}%
  \BibitemOpen
  \bibfield  {author} {\bibinfo {author} {\bibfnamefont {S.~R.~M.}\
  \bibnamefont {Bahcall J.~N.}},\ }\href {\doibase 10.1086/158905} {\bibfield
  {journal} {\bibinfo  {journal} {Astrophys. J.}\ }\textbf {\bibinfo {volume}
  {246}},\ \bibinfo {pages} {122} (\bibinfo {year} {1981})}\BibitemShut
  {NoStop}%
\bibitem [{\citenamefont {Monz}\ \emph {et~al.}(2009)\citenamefont {Monz},
  \citenamefont {Kim}, \citenamefont {Hansel}, \citenamefont {Riebe},
  \citenamefont {Villar}, \citenamefont {Schindler}, \citenamefont {Chwalla},
  \citenamefont {Hennrich},\ and\ \citenamefont
  {Blatt}}]{monz_realisation_2009}%
  \BibitemOpen
  \bibfield  {author} {\bibinfo {author} {\bibfnamefont {T.}~\bibnamefont
  {Monz}}, \bibinfo {author} {\bibfnamefont {K.}~\bibnamefont {Kim}}, \bibinfo
  {author} {\bibfnamefont {W.}~\bibnamefont {Hansel}}, \bibinfo {author}
  {\bibfnamefont {M.}~\bibnamefont {Riebe}}, \bibinfo {author} {\bibfnamefont
  {A.~S.}\ \bibnamefont {Villar}}, \bibinfo {author} {\bibfnamefont
  {P.}~\bibnamefont {Schindler}}, \bibinfo {author} {\bibfnamefont
  {M.}~\bibnamefont {Chwalla}}, \bibinfo {author} {\bibfnamefont
  {M.}~\bibnamefont {Hennrich}}, \ and\ \bibinfo {author} {\bibfnamefont
  {R.}~\bibnamefont {Blatt}},\ }\href {\doibase 10.1103/PhysRevLett.102.040501}
  {\bibfield  {journal} {\bibinfo  {journal} {Physical Review Letters}\
  }\textbf {\bibinfo {volume} {102}},\ \bibinfo {pages} {040501} (\bibinfo
  {year} {2009})}\BibitemShut {NoStop}%
\bibitem [{\citenamefont {Scott}\ and\ \citenamefont
  {Tout}(1989)}]{scott_nearest_1989}%
  \BibitemOpen
  \bibfield  {author} {\bibinfo {author} {\bibfnamefont {D.}~\bibnamefont
  {Scott}}\ and\ \bibinfo {author} {\bibfnamefont {C.~A.}\ \bibnamefont
  {Tout}},\ }\href {\doibase 10.1093/mnras/241.2.109} {\bibfield  {journal}
  {\bibinfo  {journal} {Monthly Notices of the Royal Astronomical Society}\
  }\textbf {\bibinfo {volume} {241}},\ \bibinfo {pages} {109} (\bibinfo {year}
  {1989})}\BibitemShut {NoStop}%
\bibitem [{\citenamefont {Moltchanov}(2012)}]{moltchanov_distance_2012}%
  \BibitemOpen
  \bibfield  {author} {\bibinfo {author} {\bibfnamefont {D.}~\bibnamefont
  {Moltchanov}},\ }\href {\doibase 10.1016/j.adhoc.2012.02.005} {\bibfield
  {journal} {\bibinfo  {journal} {Ad Hoc Networks}\ }\textbf {\bibinfo {volume}
  {10}},\ \bibinfo {pages} {1146 } (\bibinfo {year} {2012})}\BibitemShut
  {NoStop}%
\bibitem [{\citenamefont {Crane}()}]{crane_wolfram_2019}%
  \BibitemOpen
  \bibfield  {author} {\bibinfo {author} {\bibfnamefont {E.}~\bibnamefont
  {Crane}},\ }\href@noop {} {\bibinfo  {journal} {Wolfram Demonstrations
  Project}\ ,\ \bibinfo {pages} {in preparation}}\BibitemShut {NoStop}%
\bibitem [{\citenamefont {Fewell}(2006)}]{fewell_area_2006}%
  \BibitemOpen
\bibfield  {journal} {  }\bibfield  {author} {\bibinfo {author} {\bibfnamefont
  {M.}~\bibnamefont {Fewell}},\ }\href@noop {} {\bibfield  {journal} {\bibinfo
  {journal} {Australian Government, Department of Defense Technical Report
  DSTO-TN-0722}\ } (\bibinfo {year} {2006})}\BibitemShut {NoStop}%
\bibitem [{\citenamefont {Auerbach}(2012)}]{auerbach_interacting_2012}%
  \BibitemOpen
  \bibfield  {author} {\bibinfo {author} {\bibfnamefont {A.}~\bibnamefont
  {Auerbach}},\ }\href@noop {} {\emph {\bibinfo {title} {Interacting Electrons
  and Quantum Magnetism}}},\ Graduate Texts in Contemporary Physics\ (\bibinfo
  {publisher} {Springer New York},\ \bibinfo {year} {2012})\BibitemShut
  {NoStop}%
\bibitem [{\citenamefont {Kubo}\ and\ \citenamefont
  {Uchinami}(1975)}]{kubo_antiferromagnetic_1975}%
  \BibitemOpen
  \bibfield  {author} {\bibinfo {author} {\bibfnamefont {K.}~\bibnamefont
  {Kubo}}\ and\ \bibinfo {author} {\bibfnamefont {M.}~\bibnamefont
  {Uchinami}},\ }\href {\doibase 10.1143/PTP.54.1289} {\bibfield  {journal}
  {\bibinfo  {journal} {Progress of Theoretical Physics}\ }\textbf {\bibinfo
  {volume} {54}},\ \bibinfo {pages} {1289} (\bibinfo {year}
  {1975})}\BibitemShut {NoStop}%
\bibitem [{\citenamefont {Polatsek}\ and\ \citenamefont
  {Becker}(1996)}]{polatsek_ground-state_1996}%
  \BibitemOpen
  \bibfield  {author} {\bibinfo {author} {\bibfnamefont {G.}~\bibnamefont
  {Polatsek}}\ and\ \bibinfo {author} {\bibfnamefont {K.~W.}\ \bibnamefont
  {Becker}},\ }\href {\doibase 10.1103/PhysRevB.54.1637} {\bibfield  {journal}
  {\bibinfo  {journal} {Physical Review B}\ }\textbf {\bibinfo {volume} {54}},\
  \bibinfo {pages} {1637} (\bibinfo {year} {1996})}\BibitemShut {NoStop}%
\bibitem [{\citenamefont {Crane}\ \emph {et~al.}(2018)\citenamefont {Crane},
  \citenamefont {Kolker}, \citenamefont {Stock}, \citenamefont {Stavrias},
  \citenamefont {Saeedi}, \citenamefont {van Loon}, \citenamefont {Murdin},\
  and\ \citenamefont {Curson}}]{crane_fabrication_2018}%
  \BibitemOpen
  \bibfield  {author} {\bibinfo {author} {\bibfnamefont {E.}~\bibnamefont
  {Crane}}, \bibinfo {author} {\bibfnamefont {A.}~\bibnamefont {Kolker}},
  \bibinfo {author} {\bibfnamefont {T.~J.~Z.}\ \bibnamefont {Stock}}, \bibinfo
  {author} {\bibfnamefont {N.}~\bibnamefont {Stavrias}}, \bibinfo {author}
  {\bibfnamefont {K.}~\bibnamefont {Saeedi}}, \bibinfo {author} {\bibfnamefont
  {M.~A.~W.}\ \bibnamefont {van Loon}}, \bibinfo {author} {\bibfnamefont
  {B.~M.}\ \bibnamefont {Murdin}}, \ and\ \bibinfo {author} {\bibfnamefont
  {N.~J.}\ \bibnamefont {Curson}},\ }\href
  {https://doi.org/10.1088/1742-6596/1079/1/012010} {\bibfield  {journal}
  {\bibinfo  {journal} {Journal of Physics Conference Series}\ }\textbf
  {\bibinfo {volume} {1079}},\ \bibinfo {pages} {012010} (\bibinfo {year}
  {2018})}\BibitemShut {NoStop}%
\bibitem [{\citenamefont {Portis}\ \emph {et~al.}(1953)\citenamefont {Portis},
  \citenamefont {Kip}, \citenamefont {Kittel},\ and\ \citenamefont
  {Brattain}}]{portis_electron_1953}%
  \BibitemOpen
  \bibfield  {author} {\bibinfo {author} {\bibfnamefont {A.~M.}\ \bibnamefont
  {Portis}}, \bibinfo {author} {\bibfnamefont {A.~F.}\ \bibnamefont {Kip}},
  \bibinfo {author} {\bibfnamefont {C.}~\bibnamefont {Kittel}}, \ and\ \bibinfo
  {author} {\bibfnamefont {W.~H.}\ \bibnamefont {Brattain}},\ }\href {\doibase
  10.1103/PhysRev.90.988} {\bibfield  {journal} {\bibinfo  {journal} {Physical
  Review}\ }\textbf {\bibinfo {volume} {90}},\ \bibinfo {pages} {988} (\bibinfo
  {year} {1953})}\BibitemShut {NoStop}%
\bibitem [{\citenamefont {Pi\~neiro Orioli}\ \emph {et~al.}(2018)\citenamefont
  {Pi\~neiro Orioli}, \citenamefont {Signoles}, \citenamefont {Wildhagen},
  \citenamefont {G\"unter}, \citenamefont {Berges}, \citenamefont {Whitlock},\
  and\ \citenamefont {Weidem\"uller}}]{signoles_relaxation_2017}%
  \BibitemOpen
  \bibfield  {author} {\bibinfo {author} {\bibfnamefont {A.}~\bibnamefont
  {Pi\~neiro Orioli}}, \bibinfo {author} {\bibfnamefont {A.}~\bibnamefont
  {Signoles}}, \bibinfo {author} {\bibfnamefont {H.}~\bibnamefont {Wildhagen}},
  \bibinfo {author} {\bibfnamefont {G.}~\bibnamefont {G\"unter}}, \bibinfo
  {author} {\bibfnamefont {J.}~\bibnamefont {Berges}}, \bibinfo {author}
  {\bibfnamefont {S.}~\bibnamefont {Whitlock}}, \ and\ \bibinfo {author}
  {\bibfnamefont {M.}~\bibnamefont {Weidem\"uller}},\ }\href {\doibase
  10.1103/PhysRevLett.120.063601} {\bibfield  {journal} {\bibinfo  {journal}
  {Phys. Rev. Lett.}\ }\textbf {\bibinfo {volume} {120}},\ \bibinfo {pages}
  {063601} (\bibinfo {year} {2018})}\BibitemShut {NoStop}%
\bibitem [{\citenamefont {Schuckert}\ \emph {et~al.}(2018)\citenamefont
  {Schuckert}, \citenamefont {Pi\~neiro Orioli},\ and\ \citenamefont
  {Berges}}]{schuckert_nonequilibrium_2018}%
  \BibitemOpen
  \bibfield  {author} {\bibinfo {author} {\bibfnamefont {A.}~\bibnamefont
  {Schuckert}}, \bibinfo {author} {\bibfnamefont {A.}~\bibnamefont {Pi\~neiro
  Orioli}}, \ and\ \bibinfo {author} {\bibfnamefont {J.}~\bibnamefont
  {Berges}},\ }\href {\doibase 10.1103/PhysRevB.98.224304} {\bibfield
  {journal} {\bibinfo  {journal} {Phys. Rev. B}\ }\textbf {\bibinfo {volume}
  {98}},\ \bibinfo {pages} {224304} (\bibinfo {year} {2018})}\BibitemShut
  {NoStop}%
\bibitem [{\citenamefont {Ambegaokar}\ and\ \citenamefont
  {Troyer}(2010)}]{ambegaokar_estimating_10}%
  \BibitemOpen
  \bibfield  {author} {\bibinfo {author} {\bibfnamefont {V.}~\bibnamefont
  {Ambegaokar}}\ and\ \bibinfo {author} {\bibfnamefont {M.}~\bibnamefont
  {Troyer}},\ }\href {\doibase 10.1119/1.3247985} {\bibfield  {journal}
  {\bibinfo  {journal} {American Journal of Physics}\ }\textbf {\bibinfo
  {volume} {78}},\ \bibinfo {pages} {150} (\bibinfo {year} {2010})}\BibitemShut
  {NoStop}%
\bibitem [{\citenamefont {Saffman}\ \emph {et~al.}(2010)\citenamefont
  {Saffman}, \citenamefont {Walker},\ and\ \citenamefont
  {M\o{}lmer}}]{RevModPhys.82.2313}%
  \BibitemOpen
  \bibfield  {author} {\bibinfo {author} {\bibfnamefont {M.}~\bibnamefont
  {Saffman}}, \bibinfo {author} {\bibfnamefont {T.~G.}\ \bibnamefont {Walker}},
  \ and\ \bibinfo {author} {\bibfnamefont {K.}~\bibnamefont {M\o{}lmer}},\
  }\href {\doibase 10.1103/RevModPhys.82.2313} {\bibfield  {journal} {\bibinfo
  {journal} {Rev. Mod. Phys.}\ }\textbf {\bibinfo {volume} {82}},\ \bibinfo
  {pages} {2313} (\bibinfo {year} {2010})}\BibitemShut {NoStop}%
\bibitem [{\citenamefont {Kohn}\ and\ \citenamefont
  {Luttinger}(1955)}]{kohn_theory_1955}%
  \BibitemOpen
  \bibfield  {author} {\bibinfo {author} {\bibfnamefont {W.}~\bibnamefont
  {Kohn}}\ and\ \bibinfo {author} {\bibfnamefont {J.~M.}\ \bibnamefont
  {Luttinger}},\ }\href {\doibase 10.1103/PhysRev.98.915} {\bibfield  {journal}
  {\bibinfo  {journal} {Phys. Rev.}\ }\textbf {\bibinfo {volume} {98}},\
  \bibinfo {pages} {915} (\bibinfo {year} {1955})}\BibitemShut {NoStop}%
\bibitem [{\citenamefont {Thomas}\ \emph {et~al.}(1981)\citenamefont {Thomas},
  \citenamefont {Capizzi}, \citenamefont {DeRosa}, \citenamefont {Bhatt},\ and\
  \citenamefont {Rice}}]{thomas_optical_1981}%
  \BibitemOpen
  \bibfield  {author} {\bibinfo {author} {\bibfnamefont {G.~A.}\ \bibnamefont
  {Thomas}}, \bibinfo {author} {\bibfnamefont {M.}~\bibnamefont {Capizzi}},
  \bibinfo {author} {\bibfnamefont {F.}~\bibnamefont {DeRosa}}, \bibinfo
  {author} {\bibfnamefont {R.~N.}\ \bibnamefont {Bhatt}}, \ and\ \bibinfo
  {author} {\bibfnamefont {T.~M.}\ \bibnamefont {Rice}},\ }\href {\doibase
  10.1103/PhysRevB.23.5472} {\bibfield  {journal} {\bibinfo  {journal}
  {Physical Review B}\ }\textbf {\bibinfo {volume} {23}},\ \bibinfo {pages}
  {5472} (\bibinfo {year} {1981})}\BibitemShut {NoStop}%
\bibitem [{\citenamefont {Ramdas}\ and\ \citenamefont
  {Rodriguez}(1981)}]{ramdas_spectroscopy_1981}%
  \BibitemOpen
  \bibfield  {author} {\bibinfo {author} {\bibfnamefont {A.~K.}\ \bibnamefont
  {Ramdas}}\ and\ \bibinfo {author} {\bibfnamefont {S.}~\bibnamefont
  {Rodriguez}},\ }\href {\doibase 10.1088/0034-4885/44/12/002} {\bibfield
  {journal} {\bibinfo  {journal} {Reports on Progress in Physics}\ }\textbf
  {\bibinfo {volume} {44}},\ \bibinfo {pages} {1297} (\bibinfo {year}
  {1981})}\BibitemShut {NoStop}%
\bibitem [{\citenamefont {Li}\ \emph {et~al.}(2018)\citenamefont {Li},
  \citenamefont {Le}, \citenamefont {Litvinenko}, \citenamefont {Clowes},
  \citenamefont {Engelkamp}, \citenamefont {Pavlov}, \citenamefont {H\"ubers},
  \citenamefont {Shuman}, \citenamefont {Portsel}, \citenamefont {Lodygin},
  \citenamefont {Astrov}, \citenamefont {Abrosimov}, \citenamefont {Pidgeon},
  \citenamefont {Fisher}, \citenamefont {Zeng}, \citenamefont {Niquet},\ and\
  \citenamefont {Murdin}}]{li_radii_2018}%
  \BibitemOpen
  \bibfield  {author} {\bibinfo {author} {\bibfnamefont {J.}~\bibnamefont
  {Li}}, \bibinfo {author} {\bibfnamefont {N.~H.}\ \bibnamefont {Le}}, \bibinfo
  {author} {\bibfnamefont {K.~L.}\ \bibnamefont {Litvinenko}}, \bibinfo
  {author} {\bibfnamefont {S.~K.}\ \bibnamefont {Clowes}}, \bibinfo {author}
  {\bibfnamefont {H.}~\bibnamefont {Engelkamp}}, \bibinfo {author}
  {\bibfnamefont {S.~G.}\ \bibnamefont {Pavlov}}, \bibinfo {author}
  {\bibfnamefont {H.-W.}\ \bibnamefont {H\"ubers}}, \bibinfo {author}
  {\bibfnamefont {V.~B.}\ \bibnamefont {Shuman}}, \bibinfo {author}
  {\bibfnamefont {L.}~\bibnamefont {Portsel}}, \bibinfo {author} {\bibfnamefont
  {N.}~\bibnamefont {Lodygin}}, \bibinfo {author} {\bibfnamefont {Y.~A.}\
  \bibnamefont {Astrov}}, \bibinfo {author} {\bibfnamefont {N.~V.}\
  \bibnamefont {Abrosimov}}, \bibinfo {author} {\bibfnamefont {C.~R.}\
  \bibnamefont {Pidgeon}}, \bibinfo {author} {\bibfnamefont {A.}~\bibnamefont
  {Fisher}}, \bibinfo {author} {\bibfnamefont {Z.}~\bibnamefont {Zeng}},
  \bibinfo {author} {\bibfnamefont {Y.-M.}\ \bibnamefont {Niquet}}, \ and\
  \bibinfo {author} {\bibfnamefont {B.~N.}\ \bibnamefont {Murdin}},\ }\href
  {\doibase 10.1103/PhysRevB.98.085423} {\bibfield  {journal} {\bibinfo
  {journal} {Phys. Rev. B}\ }\textbf {\bibinfo {volume} {98}},\ \bibinfo
  {pages} {085423} (\bibinfo {year} {2018})}\BibitemShut {NoStop}%
\bibitem [{\citenamefont {Chick}\ \emph {et~al.}(2017)\citenamefont {Chick},
  \citenamefont {Stavrias}, \citenamefont {Saeedi}, \citenamefont {Redlich},
  \citenamefont {Greenland}, \citenamefont {Matmon}, \citenamefont {Naftaly},
  \citenamefont {Pidgeon}, \citenamefont {Aeppli},\ and\ \citenamefont
  {Murdin}}]{chick_coherent_2017}%
  \BibitemOpen
  \bibfield  {author} {\bibinfo {author} {\bibfnamefont {S.}~\bibnamefont
  {Chick}}, \bibinfo {author} {\bibfnamefont {N.}~\bibnamefont {Stavrias}},
  \bibinfo {author} {\bibfnamefont {K.}~\bibnamefont {Saeedi}}, \bibinfo
  {author} {\bibfnamefont {B.}~\bibnamefont {Redlich}}, \bibinfo {author}
  {\bibfnamefont {P.~T.}\ \bibnamefont {Greenland}}, \bibinfo {author}
  {\bibfnamefont {G.}~\bibnamefont {Matmon}}, \bibinfo {author} {\bibfnamefont
  {M.}~\bibnamefont {Naftaly}}, \bibinfo {author} {\bibfnamefont {C.~R.}\
  \bibnamefont {Pidgeon}}, \bibinfo {author} {\bibfnamefont {G.}~\bibnamefont
  {Aeppli}}, \ and\ \bibinfo {author} {\bibfnamefont {B.~N.}\ \bibnamefont
  {Murdin}},\ }\href {\doibase 10.1038/ncomms16038} {\bibfield  {journal}
  {\bibinfo  {journal} {Nature Communications}\ }\textbf {\bibinfo {volume}
  {8}},\ \bibinfo {pages} {16038} (\bibinfo {year} {2017})}\BibitemShut
  {NoStop}%
\bibitem [{\citenamefont {Hahn}(2005)}]{hahn_vegas_2005}%
  \BibitemOpen
  \bibfield  {author} {\bibinfo {author} {\bibfnamefont {T.}~\bibnamefont
  {Hahn}},\ }\href {\doibase https://doi.org/10.1016/j.cpc.2005.01.010}
  {\bibfield  {journal} {\bibinfo  {journal} {Computer Physics Communications}\
  }\textbf {\bibinfo {volume} {168}},\ \bibinfo {pages} {78} (\bibinfo {year}
  {2005})}\BibitemShut {NoStop}%
\end{thebibliography}%
\clearpage
\appendix
\section{Exchange calculation with multivalley wavefunctions}
\label{app_jex_calc}
In order to calculate the exchange interaction between two donors we use the Heitler-London approximation~\cite{koiller_exchange_2001}
\begin{align}\label{eq:HL}
J=\int d\bm{r_1} d\bm{r_2} \, &\psi_1^{*}(\bm{r}_1)\psi_2^{*}(\bm{r}_2-\bm{R})\frac{e^2}{4\pi\epsilon_0\epsilon_r|\bm{r}_1-\bm{r}_2|} \notag \\&\times\psi_2(\bm{r}_1-\bm{R})\psi_1(\bm{r}_2),
\end{align}
where $\bm{R}=\bm{R}_2-\bm{R}_1$ is the separation vector between the two donors and $\epsilon_r=11.4$ is the dielectric constant of silicon.
The wavefuction of each donor can be either the $1s\text{A}_1$ ground state, the $2p_0$ or $2p_{\pm}$ excited states. In the multivalley effective mass theory all of these wavefunctions can be expanded as
\begin{equation}\label{eq:mv}
\psi^{(j)}(\bm{r})=\sum_{\mu}F^{(j)}_{\mu}(\bm{r})\phi_{\mu}(\bm{r}),
\end{equation}
where $j=1,2,3$ indicates the $1s\text{A}_1$ ground state, the $2p_0$ and $2p_{\pm}$ excited states, respectively; $\mu=\pm x,\pm y,\pm z$ indicates the valleys of silicon's conduction band minima, $\phi_{\mu}(\bm{r})=e^{i\bm{k}_{\mu}.\bm{r}}u_{\mu}(\bm{r})$ are the Bloch functions at the minima, $|\bm{k}_{\mu}|=0.84\times 2\pi/a_0$ where $a_0=0.543$nm is the lattice constant. The $F_{\mu}(\bm{r})$ are the envelope functions. For the $1s\text{A}_1$ state the envelope function of the $+z$ valley is~\cite{kohn_theory_1955}
\begin{equation}
F^{(1)}_{+z}(\bm{r})=\frac{\exp\left[-\left(\frac{x^2+y^2}{(\alpha a_1)^2}+\frac{z^2}{(\alpha b_1)^2}\right)^{1/2}\right]}{\sqrt{6 \pi (\alpha a_1)^2 (\alpha b_1)}},
\end{equation}
where $a_1=2.42$nm and $b_1=1.39$ nm. The factor $\alpha$ accounts for the contraction of the ground state~\cite{thomas_optical_1981} due to the central cell correction (CCC)  and is given by $\alpha=\sqrt{E_{\text{SV}}/E_B}$ where $E_{\text{SV}}=29.7$ meV is the binding energy obtained from a single valley theory without the CCC, as done in Ref.~\cite{kohn_theory_1955},  and $E_B$ is the experimental binding energy which is $45.58$meV for Si:P and $53.77$meV for Si:As~\cite{ramdas_spectroscopy_1981}. The other envelope functions are obtained by using $F_{\mu}=F_{-\mu}$ and cyclic permutations of $x,y,z$. 

The excited state energies and wavefunctions of Si:P are identical to those of Si:As~\cite{ramdas_spectroscopy_1981,li_radii_2018}, and is dependent on the polarization of the light field~\cite{chick_coherent_2017}. For polarization with the unit vector $\bm{\epsilon}=[\epsilon_x,\epsilon_y,\epsilon_z]$ the excited states are also given by Eq.~\eqref{eq:mv} but the envelope functions are now
\begin{align}
F^{(2)}_{+z}(\bm{r}&=\frac{\epsilon_z z}{\sqrt{2\pi a_2^2 b_2^3}}  \exp\left[-\sqrt{\frac{x^2+y^2}{a_2^2}+\frac{z^2}{b_2^2}}\right], \nonumber \\
F^{(3)}_{+z}(\bm{r}&=\frac{\epsilon_x x+\epsilon_y y}{\sqrt{4\pi a_3^4 b_3}} \exp\left[-\sqrt{\frac{x^2+y^2}{a_3^2}+\frac{z^2}{b_3^2}}\right],
\end{align}
where $a_2=3.68$nm, $b_2=2.23$nm, $a_3=5.45$nm, and $b_3=3.35$nm. The other envelope functions can again be derived using $F_{\mu}=F_{-\mu}$ and cyclic permutations of $x,y,z$. 

We can further expand the Bloch function in Eq.~\eqref{eq:mv} in terms of plane waves as $\phi_{\mu}(\bm{r})=\sum_{\bm{G}}c_{\bm{G}}e^{i \bm{G}.\bm{r}}$ where $\bm{G}$ is the reciprocal lattice vector. Substituting this into Eq.~\eqref{eq:mv} and then Eq.~\eqref{eq:HL}, neglecting the fast oscillating terms in the resulting integrand, and using the equality $\sum_{\bm{G}}|c_{\bm{G}}|^2=1$ we arrive at
\begin{equation}
J=2\sum_{\mu,\nu}j_{\mu,\nu} \cos({\bm{k}_{\mu}.\bm{R}})\cos({\bm{k}_{\nu}.\bm{R}}),
\end{equation}
where
\begin{align}
j_{\mu,\nu}=\int d \bm{r}_1 d \bm{r}_2 \, &F^*_{1,\mu}(\bm{r}_1)F^*_{2,\nu}(\bm{r}_2-\bm{R}) \frac{e^2}{4\pi\epsilon_0\epsilon_r|r_1-r_2|} \notag \\& \times F_{2,\nu}(\bm{r}_1-\bm{R})F_{1,\mu}(\bm{r}_1).
\end{align}
These highly oscillating integrals are evaluated numerically with the Vegas Package~\cite{hahn_vegas_2005}.

\section{Monte Carlo Simulation}\label{app_simu}

\begin{algorithm}[H]
\caption{\label{alg_viabletest} Check isolation of a point by $R_v$(squares are labelled on Fig.~\ref{fig_viabletest})}
\begin{algorithmic}
\REQUIRE $D$: array containing coordinates of all points in distribution
\REQUIRE $P$: array containing coordinates of point being tested for isolation
\STATE $a \leftarrow$ $R_v \cos{45}$
\STATE i $\leftarrow 0$
\STATE Mark all elements of D as viable
\WHILE{$i<$Length($D$)}
\IF{P is marked as viable}
	\IF{$P_x - a < D[i]_x < P_x + a$}
		\IF{$P_y - a < D[i]_y < P_y + a$}
			\STATE  Mark $D[i]$ and P as non-viable
			\STATE $i=i+1$
		\ENDIF	
	\ENDIF
	\IF{$P_x - R_v < D[i]_x < P_x + R_v$}
		\IF{$P_y - R_v < D[i]_y < P_y + R_v$}
			\STATE  $b \leftarrow Sqrt((D[i]_x - P_x)^2 + (D[i]_y- P_y)^2)$
			\IF{$b < R_v$}
				\STATE Mark $D[i]$ and P as non-viable
				\STATE $i=i+1$
			\ENDIF
		\ENDIF	
	\ENDIF
	\STATE $i=i+1$
\ENDIF
\ENDWHILE
\end{algorithmic}
\end{algorithm}

\begin{figure}
\includegraphics{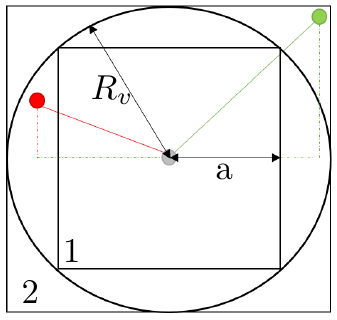}
\caption{\label{fig_viabletest} \textbf{Monte-Carlo simulation isolation checking algorithm described in Alg. \ref{alg_viabletest}}.}
\end{figure}

The Monte-Carlo simulation was written in php with a MySQLi database.  The front-end is written in html/css so that simulations can be run through a browser and made available on the internet. It scales optimally with the number of dopants, i.e. O(n\sub{c} + n\sub{r}), where n\sub{c} is the number of control dopants and n\sub{r} is the number of readout dopants, such that it allows for computations on large densities and such that the wait time is minimal.  Optimal scaling stems principally from the avoidance of full Pythagoras computations. This is achieved with four different techniques.  As regards the first step in the whole algorithm which is to analyse the controls in order to identify all viable controls, the avoidance of Pythagoras is achieved by first partitioning the space according to the expected mean nearest neighbour distance, second by using Pythagoras only for points lying between the enclosing squares (Fig.~\ref{fig_viabletest}), third by skipping to the next point as soon as conditions are found to be breached and fourth by reciprocating information between adjacent points.

\begin{figure}
\includegraphics{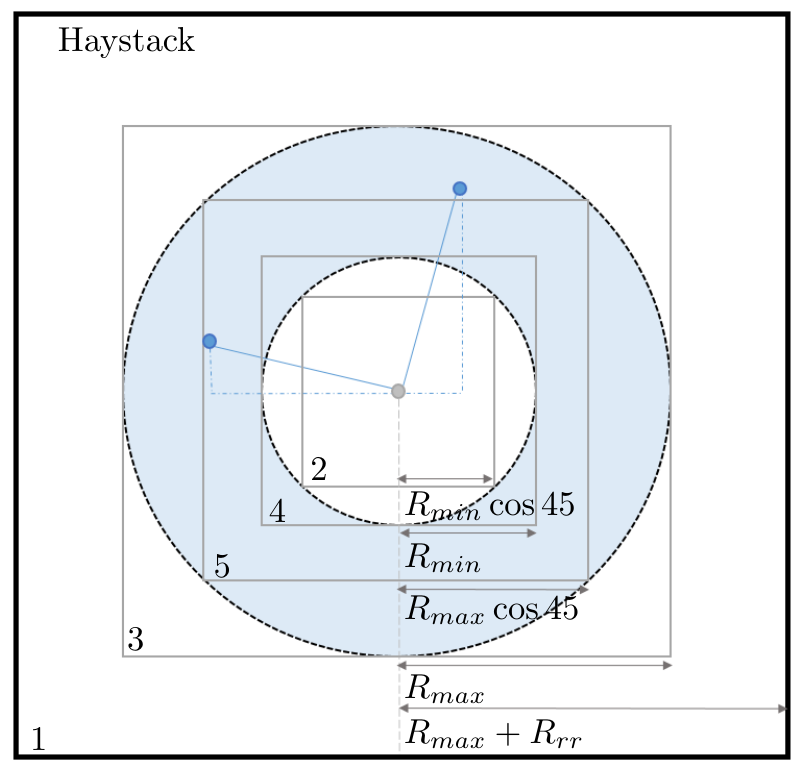}
\caption{\label{fig_haystack}\textbf{Schematic explanation of the Haystack algorithm described in Alg. \ref{alg_haystack}}. In the case of there being many readouts in one shell, this method of recording reciprocal non-viability cuts calculation time approximately in half.}
\end{figure}

As regards the second step which is to work within the viable control distribution and the full readout distribution, avoidance of Pythagoras is achieved by using the search sequence represented diagrammatically in Fig.~\ref{fig_haystack}, referred to as a Haystack search. 

\begin{algorithm}[H]
\caption{\label{alg_haystack}Find which controls and readouts are active (squares are labelled on Fig.~\ref{fig_haystack})}
\begin{algorithmic}
\REQUIRE $C_v$: array containing all viable controls (isolated by $R_{cc}$ within the control distribution: the output of Alg.~\ref{alg_viabletest})
\REQUIRE Mark all elements of $C$ as active
\REQUIRE $R$: array containing all readouts
\REQUIRE Mark all elements of $R$ as non-active
\STATE i $\leftarrow 0$
\WHILE{$i<$Length($C_v$)}
\STATE array $H \leftarrow$ elements of $R$ located within square 1 of $C_v[i]$
\IF{$H$ does not contain elements within square 3}
	\STATE Mark $C_v[i]$ as non-active
	\STATE $i=i+1$
\ELSE
	\IF{$H$ contains any element within square 2}
		\STATE Mark $C_v[i]$ as non-active
		\STATE $i=i+1$
	\ELSE
		\FOR{k indexing elements of H within square 4}
			\IF{$R[k]$ within $R_{min}$ (Pythagoras calculation)}
				\STATE Mark $C_v[i]$ as non-active
				\STATE $i=i+1$
			\ELSE
				\STATE Check $R[k]$ for isolation within $H$ (Alg. \ref{alg_viabletest}) \& if isolated, mark $R[k]$ as active. If active, array $R_a$ $\leftarrow +\: R[k]$ 
			\ENDIF
		\ENDFOR
		\FOR{k indexing of elements within square 5 and not within square 4}
			\STATE Check $R[k]$ for isolation within $H$ (Alg. \ref{alg_viabletest}) \& if isolated, mark $R[k]$ as active. If active, array $R_a$ $\leftarrow +\: R[k]$
		\ENDFOR
		\FOR{k indexing elements of H between squares 3 and 5}
			\IF{$R[k]$ within $R_{max}$ (Pythagoras calculation)}
				\STATE Check $R[k]$ for isolation within $H$ (Alg. \ref{alg_viabletest}) \& if isolated, mark $R[k]$ as active. If active, array $R_a$ $\leftarrow +\: R[k]$ 
			\ENDIF
		\ENDFOR
		\IF{Size($R_a$) $> 0$}
			\STATE Mark $C_v[i]$ as having Length($R_a$) readouts
			\STATE $i=i+1$
		\ENDIF
	\ENDIF
\ENDIF
\ENDWHILE
\end{algorithmic}
\end{algorithm}

Only the readouts close enough to a viable control and which might therefore constitute or interfere with an entangling gate are analysed. Consequently slightly more than half the readout dopants are never involved in the algorithm.

\end{document}